\documentclass[aps,epsfig,prl,twocolumn,superscriptaddress,showpacs,showkeys,email]{revtex4}
\usepackage{graphics}
\begin{document} 

\title{The {G\={o}} model revisited: Native structure and the geometric 
coupling between local and long-range contacts}

\author{Patr\'\i cia F. N. Fa\'\i sca}
\email[Correspondence to PFN Fa\'\i sca, Centro de F\'\i sica Te\'orica e
Computacional da Universidade de Lisboa, Av. Prof. Gama Pinto 2, 1649-003
Lisboa Codex, Portugal ]{E-mail:patnev@alf1.cii.fc.ul.pt}
\author{Margarida M. Telo da Gama}
\author{Ana Nunes}
\affiliation{Centro de F\'\i sica Te\'orica e
Computacional da Universidade de Lisboa, Av. Prof. Gama Pinto 2, 1649-003
Lisboa Codex, Portugal}

\begin{abstract}
\bf{Monte Carlo simulations show that long-range interactions play a major 
role in determining the folding rates of 48-mer three-dimensional lattice 
polymers modelled by the G\={o} potential.
For three target structures with different native geometries
we found a sharp increase in the folding time when the relative 
contribution of the long-range interactions to the native state's energy is 
decreased from $\sim 50\%$ towards zero. However, the dispersion of the 
simulated folding times depends strongly on the native geometry and G\={o} 
polymers folding to one of the target structures exhibit 
folding times spanning three orders of magnitude.
We have also found that, depending on the target geometry, a strong geometric 
coupling may exist between local and long-range contacts meaning that, 
when this coupling exists, the formation of long-range contacts is forced 
by the previous formation of local contacts. The absence 
of a strong geometric coupling leads to kinetics that are more sensitive 
to the interaction energy parameters; in this case the formation
of local contacts is not sufficient to promote the establishment of 
long-range ones when these are strongly penalized energetically, leading 
to longer folding times.} 
\end{abstract}

\maketitle

\section{I. Introduction}
In the last few years the idea that the native geometry governs the
overall folding kinetics of small (typically with less than 100 amino acids), 
single domain, two-state proteins has attracted considerable attention and 
prompted several new lines of research 
{\cite{PLAXCO1,DU,BAKER,SHAKH1,MARQUSEE,PFN1,WEIKL,KAYA,PFN2}}. 
A particularly important observation by Plaxco {\it et al.} 
{\cite{PLAXCO1,PLAXCO2}} 
revealed the existence of a strong correlation ($r=$0.92) between the  
experimental folding rates of 24 two-state folders and the so-called contact 
order parameter, CO, 
measuring the average sequence separation of contacting residue pairs in the 
native structure relative to the 
protein chain length. The connection between the CO (and in more general terms,
the native geometry) and the average range of amino acid 
interactions in the native fold has set a new ground for discussing an 
old-debated issue in the protein 
folding literature, namely, that of understanding the roles of local (i.e. 
close in space and in sequence) and long-range 
(i.e. close in space but distant in sequence) inter-residue 
interactions in the folding dynamics.  
Results obtained within the scope of this debate agree on the idea that 
long-range (LR) interactions play 
an important role in stabilizing the native fold 
{\cite{GO,GUTIN,BAKER1,GROMIHA1}} but there is no 
consensus on their role in the folding 
kinetics. For example, very early results obtained by G\={o} and Taketomi 
{\cite{GO}} for a 49-residue chain 
on a two-dimensional square lattice suggest that local interactions 
accelerate 
both the folding and unfolding
transitions. In Ref.{\cite{MOULT}} Unger and Moult have studied optimized 
heteropolymer sequences with chain
length $N=27$ on a three-dimensional cubic lattice and concluded that 
increasing the strength of local interactions increases the ability  
of sequences to fold. In a different study {\cite{BAKER1}}, Doyle and 
co-workers 
have found that, in the context of the Zwanzig model, the rate of folding 
increases as the contribution of the local interactions to the native 
state's energy increases. By contrast, results obtained by Abkevich
{\it{et al.}}{\cite{GUTIN}} for the Miyazawa-Jernigan lattice-polymer model 
provided evidence that, under conditions where the native state is stable, 
a 36-residue sequence on a three-dimensional cubic lattice folds to a native 
structure with mostly LR contacts two-orders of magnitude faster 
than a sequence folding to a native structure with 
predominantly local contacts. 
In Ref. {\cite{GOLDSTEIN}} Govindarajan and Goldstein have used a 
lattice model in conjunction with 
techniques drawn from the theory of spin glasses and found that optimal 
conditions 
for folding
are achieved when local interactions contribute little to the native state's 
energy.         
More recently, Gromiha and Selvaraj have analysed the 
`global' contribution of LR interactions to the folding kinetics  
by introducing a new geometrical parameter named long-range order 
(LRO){\cite{GROMIHA2}}.
The LRO, that measures the number of LR contacts in the native structure
relative to the protein chain length, was found to correlate as well as 
the CO with the folding rates of 23 (out of the 24) two-state folders 
previously studied by Plaxco {\it et al} 
{\cite{PLAXCO2}. This observation emphasizes the relative importance of LR 
interactions in protein folding kinetics.\par 
In addition, it has been shown recently that the free energy landscapes 
of single domain, two-state folders
are considerably smooth {\cite{GILLESPIE,ONUCHIC}}. This finding led to a
renewed interest {\cite{KAYA,FAN,JEWETT}}
in the G\={o} model and other modified G\={o}-type interaction schemes 
since, as for simple proteins, their energy landscapes
are relatively smooth {\cite{JEWETT}}. Indeed,  
these models do not take into account the sequence's chemical composition 
and account only for attractive interactions between native contacts 
thereby eliminating possible energetic traps. As a consequence only
geometric traps, resulting from the chain connectivity and the geometry 
of 
the native fold, will contribute to the landscape's ruggedness and thus 
G\={o} type models are said to be frustrated in a `topological' sense. 
These models are therefore particularly suited to investigate the role 
of the native state's geometry in the folding kinetics of
simple proteins.\par 
Motivated by these results, we revisit the G\={o} model to investigate 
the dependence of the
folding kinetics on LR (and local) interactions for different native 
geometries.
Our main finding is that, for G\={o}-type lattice polymers with 
$N=48$ amino acids, the LR interactions play a crucial role in 
determining the folding rates and, most importantly, this 
effect is strongly dependent on the native state's geometry. 
Indeed, we have found that, depending on
the native geometry, the dispersion of the simulated folding times spans 
up to $\approx$ 3 orders of 
magnitude when the relative strength of
LR interactions varies from zero (only local interactions contribute to 
the native state's energy) to one   
(only LR interactions contribute to the native state's energy). 
We have also found that, depending on native state's geometry, the set-up of LR 
contacts may be strongly associated  with the previous formation of local contacts. 
The existence of this geometric coupling between local and LR contacts 
explains why the observed folding kinetics may depend rather weakly on
the relative energetic contributions of local and long-range 
interactions. In proteins where this geometric coupling is stronger,
the local contacts promote the establishment of LR contacts even 
when the LR interactions are not energetically favored.\par 
The present article is organized as follows: Section II describes the model 
and methods. In Section III we present and discuss the results 
of the simulations and in Section IV we draw the conclusions.     

\section{II. Model and methods}

Protein chains with $N=48$ amino acids are modelled as self-avoiding walks 
on a three-dimensional infinite lattice. Amino acids are represented by beads 
of uniform size and 
the peptide bond, that covalently connects amino acids along the 
polypeptide chain, 
is represented by a stick of size equal to the lattice spacing. In order 
to 
mimic protein movement 
we use the so-called `kink-jump'
move set including corner-flips, end and null moves as well as 
crankshafts {\cite{BINDER}}. 
The G\={o} potential is used to model amino acid interactions which means 
that, for a 
given target native structure, equal stabilizing energies ($<0$) are 
ascribed to all the native contacts, i.e. contacts between pairs of 
beads which are present in the 
target, and neutral energies ($=0$) are ascribed to non-native 
contacts, 
i. e., contacts that are not present in the target structure.
The total energy of a conformation $\Gamma=\lbrace \vec{r_{i}} \rbrace$  
is then
given by the contact Hamiltonian
\begin{equation}
H(\lbrace \vec{r_{i}} \rbrace)=\sum_{i>j}^N
B_{ij} \Delta(\vec{r_{i}}-\vec{r_{j}}),
\label{eq:no1}
\end{equation}
where the contact function,
$\Delta (\vec{r_{i}}-\vec{r_{j}})$, is unity
if beads $i$ and $j$ form a native contact but are not covalently linked 
and is zero otherwise
and the interaction energy parameter is $B_{ij}=-\epsilon$. \par

The folding simulations follow the standard Monte Carlo (MC) Metropolis 
algorithm {\cite{METROPOLIS}}.
Each MC run starts from a randomly generated unfolded conformation 
(typically with less than 10 native contacts) and the folding dynamics is 
traced by following the evolution of the fraction of native
contacts, $Q=q/Q_{max}$, where $Q_{max}=57$ (for chains with length 
$N=48$) and $q$ is the number of native 
contacts at each MC step.
The folding time, $t$, is given by the first passage time (FPT), that is,
the number of MC steps corresponding to $Q=1.0$.

\begin{figure*}[!ht] 
{\rotatebox{270}{\resizebox{5.5cm}{7.5cm}{\includegraphics
{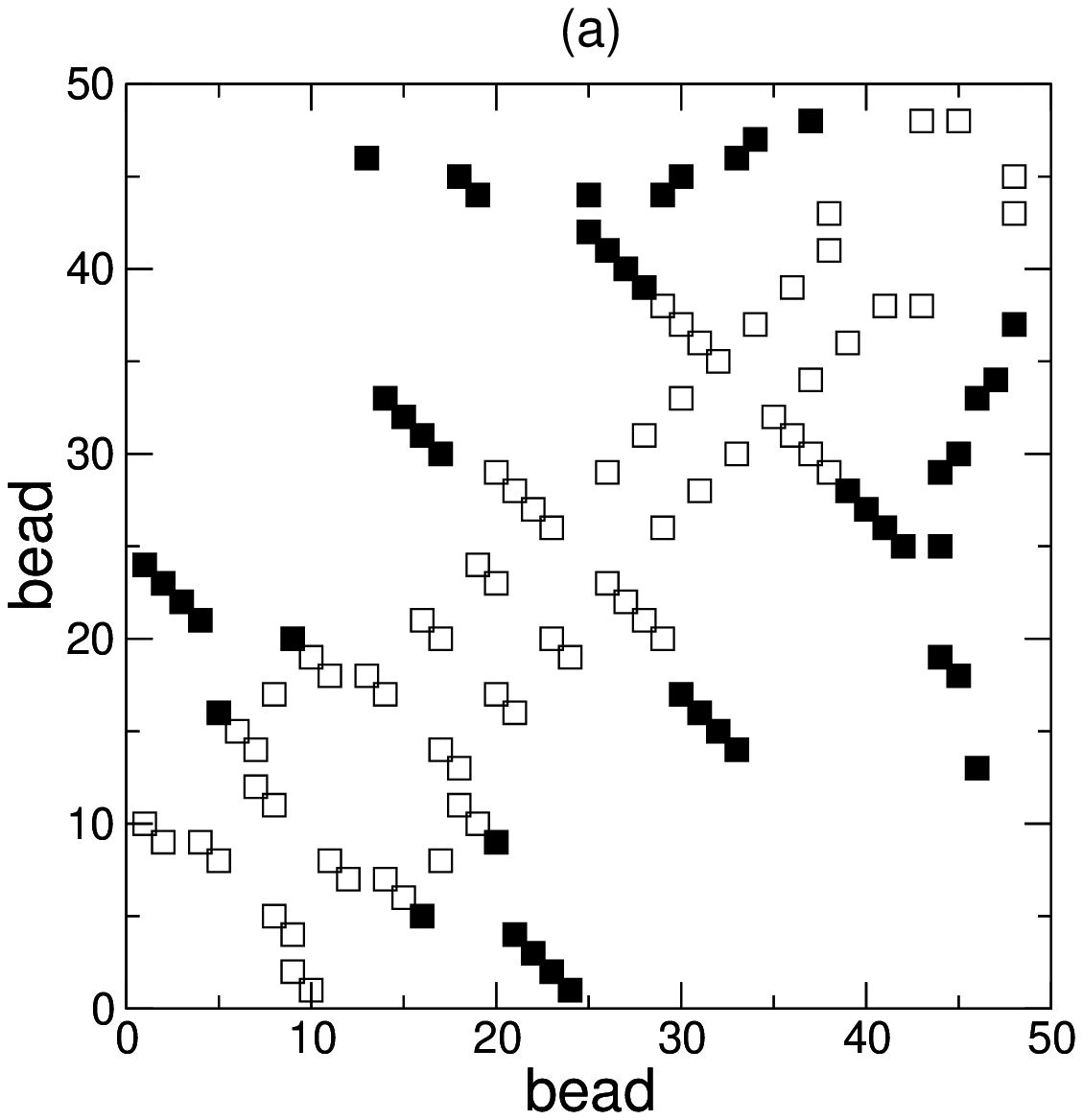}}}} 
{\rotatebox{270}{\resizebox{5.5cm}{7.5cm}{\includegraphics
{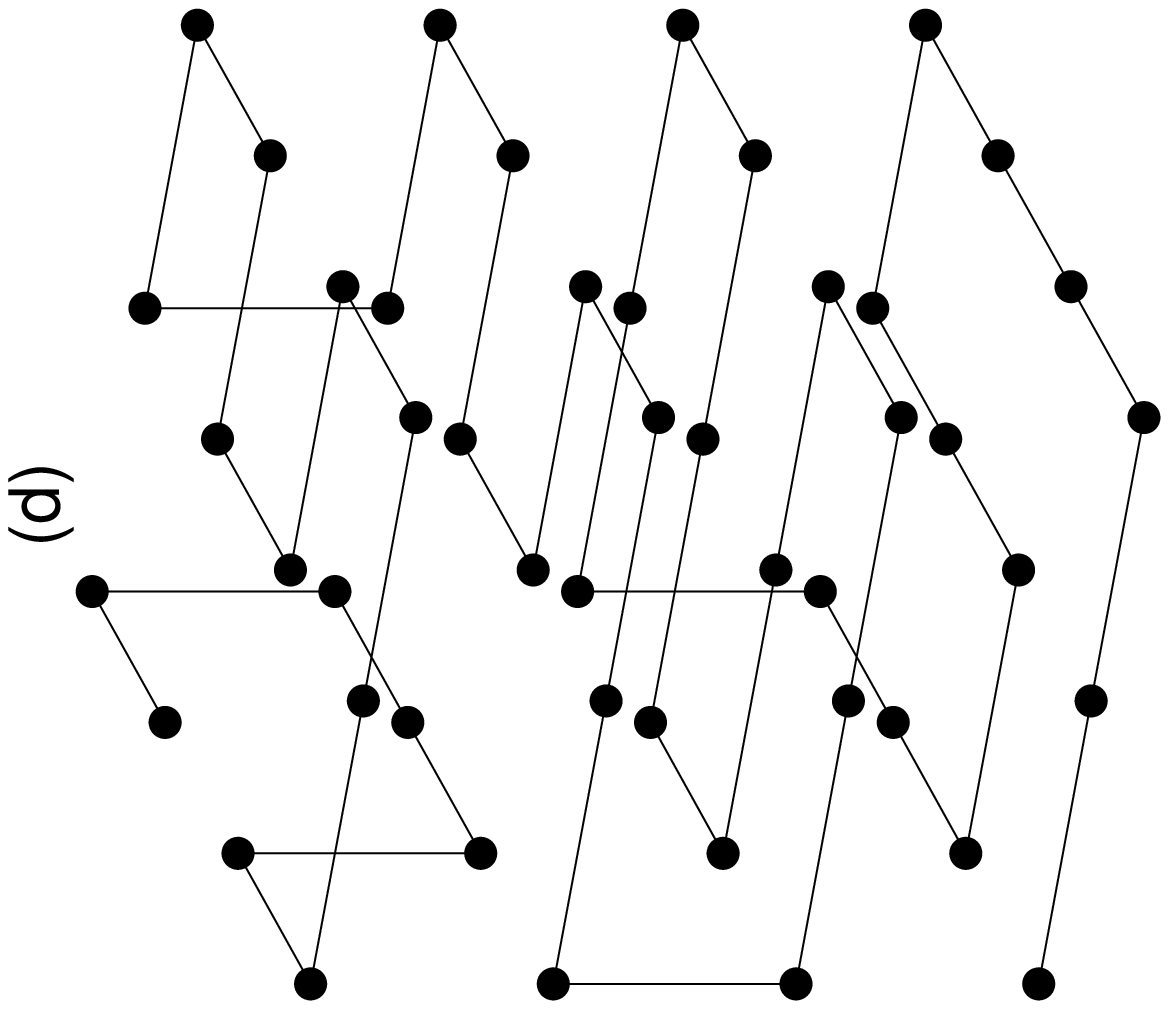}}}} \\
\vspace{0.5cm}
{\rotatebox{270}{\resizebox{5.5cm}{7.5cm}{\includegraphics
{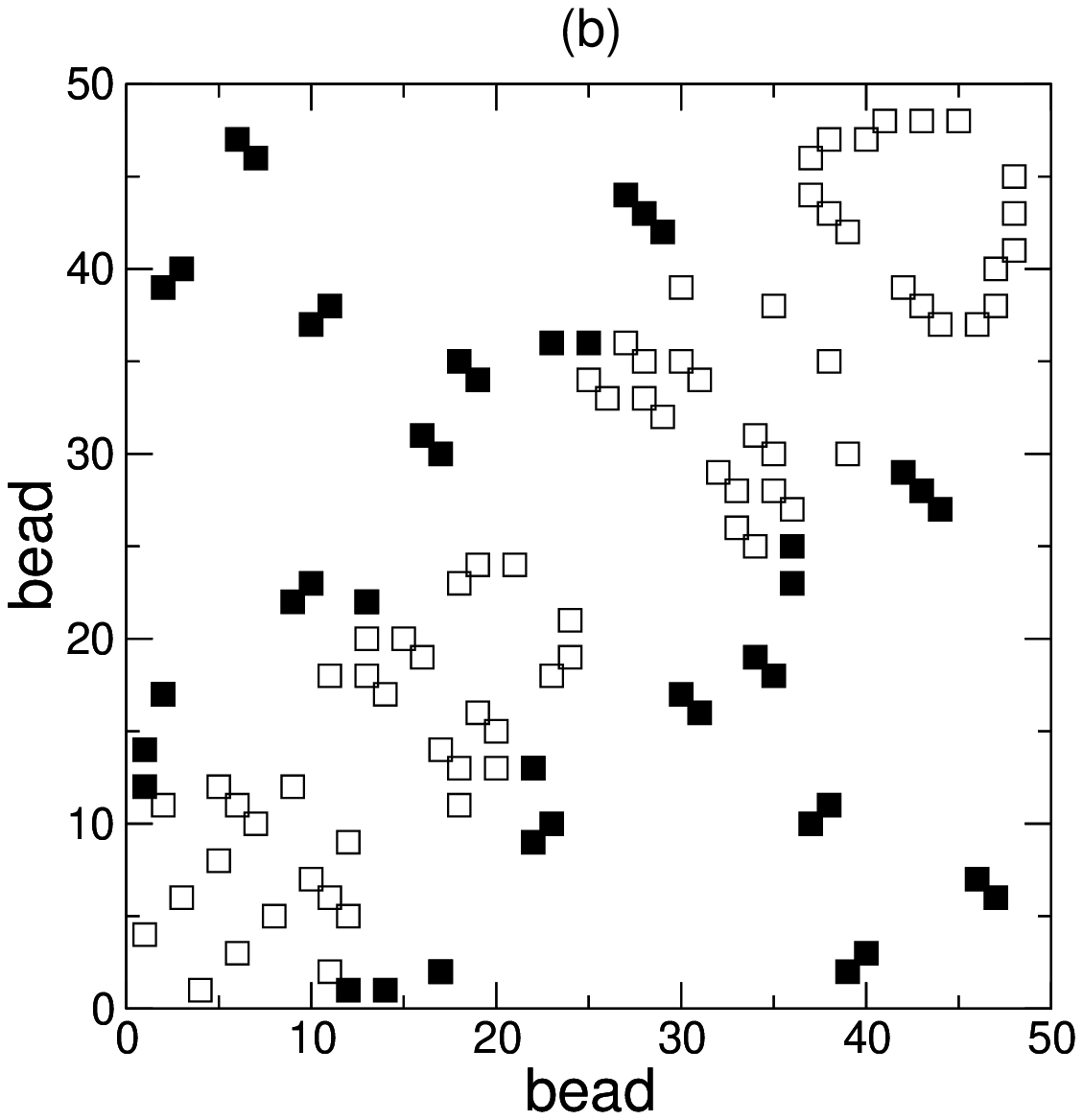}}}} \hspace{0.5cm}
{\rotatebox{270}{\resizebox{5.5cm}{7.5cm}{\includegraphics
{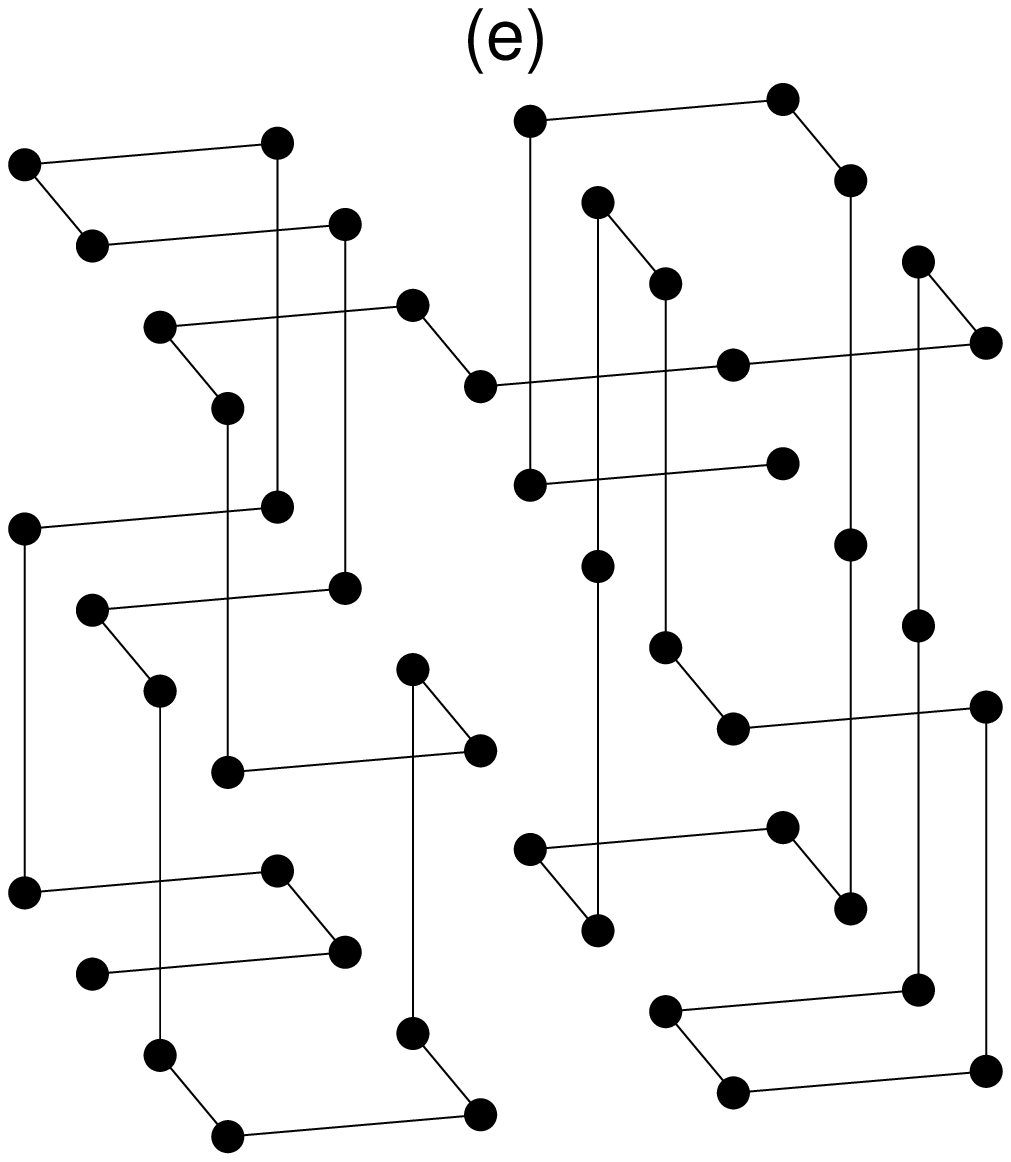}}}} \\
\vspace{1cm}
{\rotatebox{270}{\resizebox{5.5cm}{7.5cm}{\includegraphics
{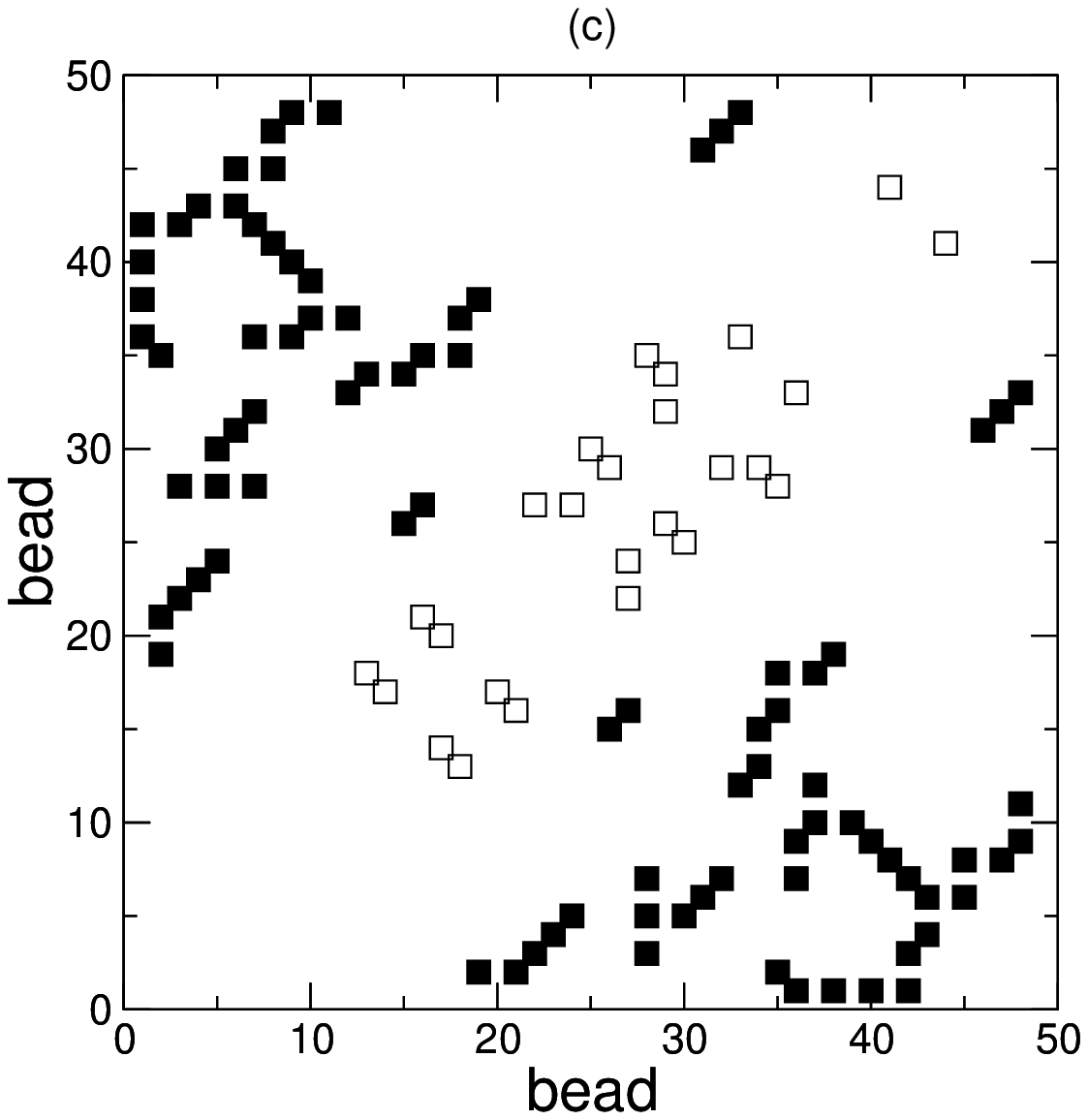}}}} \hspace{0.5cm}
{\rotatebox{270}{\resizebox{5.5cm}{7.5cm}{\includegraphics
{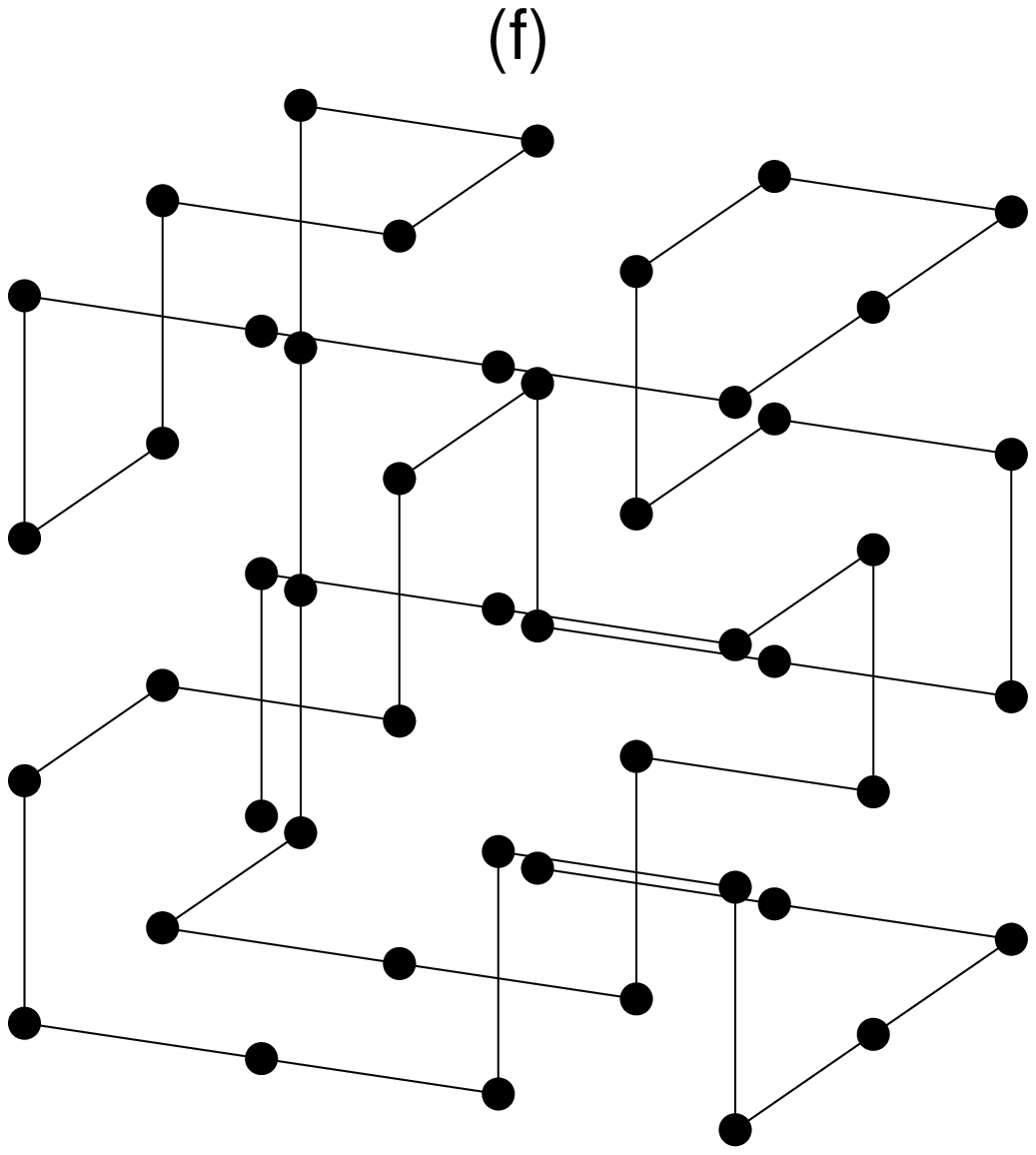}}}} 
\caption{Contact map and three-dimensional representation of structures 
$\Gamma_{1}$ (a, d), $\Gamma_{2}$ (b, e)
and $\Gamma_{3}$ (c, f). In the contact maps the black squares represent 
the long-range contacts and the white squares stand for the local contacts. 
\label{fig:no1}}
\end{figure*}

\subsection{A. Native structures}
We consider three native structures,
displaying different geometries as measured by the contact order parameter. 
These structures were found by homopolymer relaxation. In these MC simulations
a homopolymeric chain (i.e., a polymer chain composed by beads of a single 
chemical type) is launched, at temperature $T=0.7$, from a randomly generated 
conformation and relaxes, after some MC steps, to the minimum energy 
conformation. At each MC step a local random displacement of one or two beads, 
provided by the kink-jump move set, is accepted or rejected in accordance with 
the Metropolis rule. For each conformation the total energy is given by the 
contact Hamiltonian of Equation~\ref{eq:no1} where $\Delta=1$ if beads are 
in contact but not covalentely linked and is zero otherwise. The pairwise 
interaction energy parameter is $B_{ij}=-1.0$. For homopolymers of chain 
length $N=48$ on a three-dimensional cubic lattice the most stable 
conformation, evolving under the Hamiltonian of Equation~\ref{eq:no1}, is a 
cuboid with 57 contacts. Because this structure displays a maximum number of 
contacts it is generally referred to as a maximally compact structure. \par
In order to emphasize their different geometries, the low-CO (0.12) structure, 
$\Gamma_{1}$, the intermediate-CO (0.19) structure, $\Gamma_{2}$, and the 
high-CO structure (0.26) $\Gamma_{3}$, are represented in 
Figures~\ref{fig:no1}
(a)-(c) through their contact maps {\cite{CMAP}}. The corresponding 
three-dimensional structures are depicted in Figures~\ref{fig:no1}(d)-(f). 
The contact map, $C$, is an $N \times N$ matrix with entries 
$C_{ij}=1$ if beads $i$ and $j$ are in contact and zero otherwise. 
In Ref. {\cite{GROMIHA2}} Gromiha and Sevaraj have found that for real 
two-state proteins the amino acids which are close in space and separated 
by at least 10 to 15 amino acids in sequence are important determinants of 
folding rates. Motivated by this finding we define
a native contact between two beads $i$ and $j$ as a local contact if 
the backbone separation $\vert i-j \vert$ is such that 
$\vert i-j \vert \ \leq 10$ or as long-range (LR) contact if 
$\vert i-j \vert > 10$.
In Figure~\ref{fig:no1} the black squares represent the LR contacts while 
the white squares stand for the local contacts. The LRO parameter is 0.48 
for $\Gamma_{1}$, 0.44 for $\Gamma_{2}$ and 0.92 for $\Gamma_{3}$. 
We stress that the number of LR and local contacts is approximately the same
in targets $\Gamma_{1}$ and $\Gamma_{2}$.
The average LR contact length is 17.1 for $\Gamma_{1}$, 20.1 for $\Gamma_{2}$, 
and 26.5 for $\Gamma_{3}$. Table I summarizes the geometric traits of the 
three target structures.

\begin{table}
\caption{\label{}Contact order, fraction of long-range native contacts, 
$Q_{LR}$, long-range order and average long-range contact length for 
structures $\Gamma_{1}$, $\Gamma_{2}$ and $\Gamma_{3}$. $Q_{LR}$ is the 
number of LR native contacts normalised to the total number of native contacts.}
\begin{ruledtabular}
\begin{tabular}{c c c c c}
Target     &     CO    &   $Q_{LR}$  &  LRO   &   $<|i-j|_{LR}>$  \\ \hline
$\Gamma_{1}$ &    $0.129$ &     $0.40$  & $0.48$  &      $17.1$       \\
$\Gamma_{2}$ &    $0.190$ &     $0.35$  & $0.42$  &      $20.1$       \\
$\Gamma_{3}$ &    $0.259$ &     $0.77$  & $0.92$  &      $26.5$       
\end{tabular}
\end{ruledtabular}
\end{table}

\section{III. Numerical results}

\subsection{A. Simulation temperature}

The folding kinetics depends on the temperature. Indeed, when the
temperature is very high, all conformations are equally foldable and the 
kinetics becomes increasingly slower due to rapid interconversions 
(fluctuations) between unfolded states (in the high-temperature regime 
the folding time approaches the Levinthal time, i.e., it becomes exponential 
in the number of accessible conformations). On the other hand, in the 
low-temperature regime, an Arrhenius-like behaviour, characterized by 
trapping into metastable states is expected (as discussed 
in~\cite{SHAKHPRL}). Therefore, for kinetically foldable proteins, there 
must exist an intermediate temperature where the folding process is 
fastest. The existence of this 
temperature, called the optimal folding temperature, was reported in 
several studies for lattice models (sequence-specific as well as G\={o} 
models)~\cite{SHAKHPRL, JCPSHAKH,CIEPLAK,PFN3}.\par
In the present study folding kinetics is studied at the optimal folding 
temperature $T^{*}$, 
that is, the temperature that minimises the folding time, $t$. In order to 
determine
$T^{*}$ we performed MC simulations over a broad temperature range and ran a
set of 100 MC simulations at each temperature, $T$. The folding time was 
then taken as the mean FPT to the native structure averaged over the 100 MC 
runs.\par
Figure~\ref{fig:no2} reports the dependence of the folding time on the 
folding temperature for each structure and $\epsilon=0.5$. At the optimal 
folding temperature, the kinetics is not dominated by kinetic traps, and 
folding to the native state proceeds relatively fast.\par   
We stress that, at $T\geq T^{*}$, the observed
dispersion of folding times is rather small ($5.56 \pm 0.04 \le 
\log_{10}(t) \le 6.11 \pm 0.05$) and note that such behaviour is typical 
of the G\={o} and other lattice (as well as off-lattice) models 
(Ref.{\cite{KAYA}} and references therein.)\par
The functional dependence of the folding time on the temperature is 
qualitatively similar for the three structures in the high-$T$ regime. Note that in
this regime one also observes a small dispersion of the folding times.
However the reported results show that in the low-$T$ regime the 
dependence of the folding time on temperature is sensitive to the native 
state's geometry. In particular we have not observed 
folding to $\Gamma_{3}$ at low temperatures, $T < 0.28$.\par 
The optimal folding temperature, on the other hand, appears to be a 
geometry independent parameter.           
\begin{figure}
\centerline{\rotatebox{0}{\resizebox{8cm}{8cm}{\includegraphics
{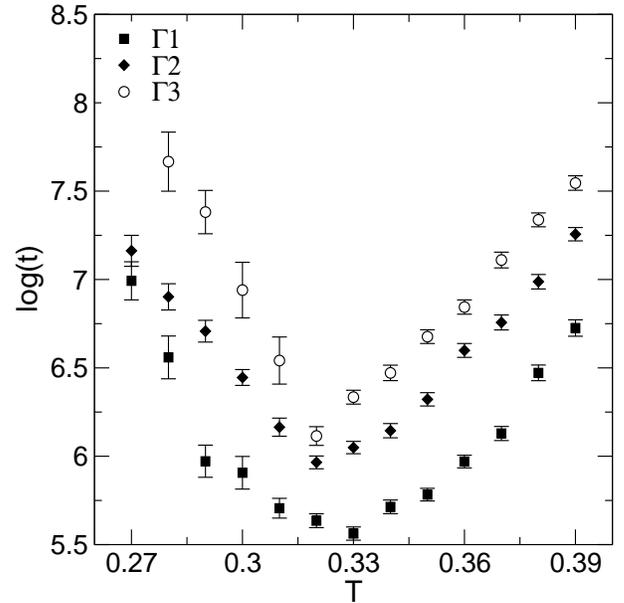}}}} 
\caption{Dependence of the logarithmic folding time, $log_{10}(t)$, on the 
simulation temperature, $T$, for structures $\Gamma_{1}$, $\Gamma_{2}$ and $\Gamma_{3}$. \label{fig:no2}}
\end{figure}

\subsection{B. Folding kinetics for different range bias}

In this section we investigate the role of LR and local contacts in the 
kinetics of
protein folding by varying the relative contributions of LR and local
interactions to the total energy in the following way: 
the energy of a conformation is given by 
\begin{equation}
H(\lbrace \vec{r_{i}} \rbrace)=\sigma H_{LR}(\lbrace \vec{r_{i}} \rbrace)+
(1-\sigma)H_{L}(\lbrace \vec{r_{i}} \rbrace),
\label{eq:no2}
\end{equation}
where the terms $H_{LR}$ and $H_{L}$ determine the overall
energy contribution of long-range and local contacts to the conformation's 
energy and are given by
\begin{equation}
H_{LR(L)}(\lbrace \vec{r_{i}} \rbrace)=-\sum_{i>j}^N \Delta_{LR(L)}(\vec{r_{i}}-\vec{r_{j}}),
\label{eq:no3}
\end{equation}
\noindent
where 
$\Delta_{LR(L)}(\vec{r_{i}}-\vec{r_{j}})$ is unity if 
beads $i$ and $j$ form a LR(local) native contact and is zero otherwise. The 
parameter $\sigma$, that we shall call range bias parameter, takes values in
$0.0 \le \sigma \le 1.0$. 
When $\sigma$ is zero all LR interactions are 
`switched-off' and only local interactions contribute
to the conformation's total energy. The reverse situation is obtained 
when $\sigma=1.0$ as in this
case only LR interactions contribute to the total energy. 
Conformation's energies given by Equation~\ref{eq:no2} 
imply that the native's state
energy varies as a function of $\sigma$. 

\begin{figure} 
\centerline{\rotatebox{0}{\resizebox{8cm}{8cm}{\includegraphics
{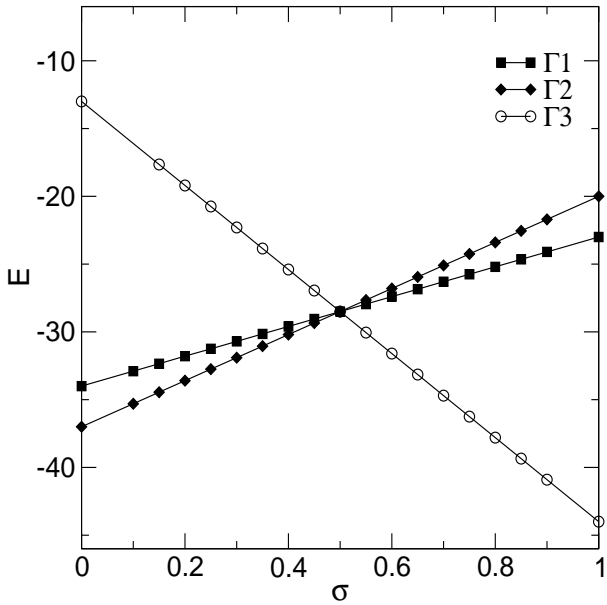}}}} 
\caption{Dependence of the native state's energy, $E$, on the range bias parameter
$\sigma$ for structures $\Gamma_{1}$, $\Gamma_{2}$ and $\Gamma_{3}$. \label{fig:no3}}
\end{figure}

Results plotted in 
Figure~\ref{fig:no3} illustrate  
the dependence of the native's state energy on the range bias parameter 
for targets $\Gamma_{1}$, $\Gamma_{2}$ and $\Gamma_{3}$. While targets 
$\Gamma_{1}$ and $\Gamma_{2}$
have predominantly local contacts and thus their energy increases 
with $\sigma$ for target $\Gamma_{3}$
the lowest native state energy is observed when $\sigma=1.0$. Since the 
native state's 
energy depends on the range bias parameter we have 
determined, for each $\sigma$,
the corresponding optimal folding temperature, $T^{*}$.\par
\begin{figure}
\centerline{\resizebox{8cm}{8cm}{\includegraphics {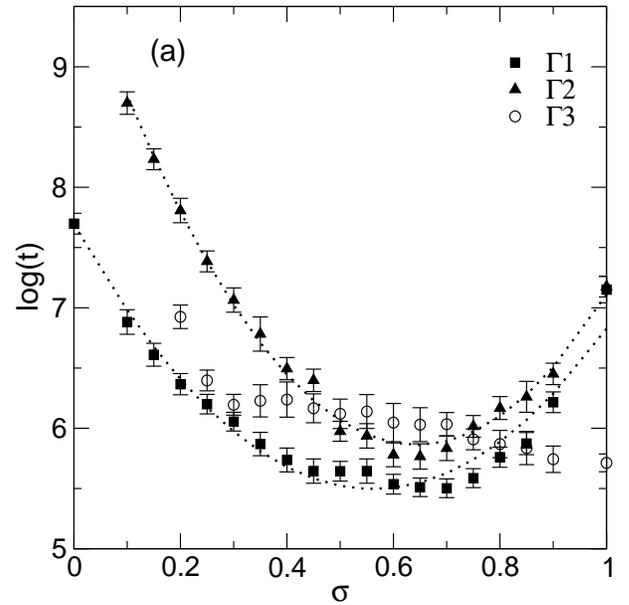}}}
\vspace{1.5cm}
\centerline{\resizebox{8cm}{8cm}{\includegraphics {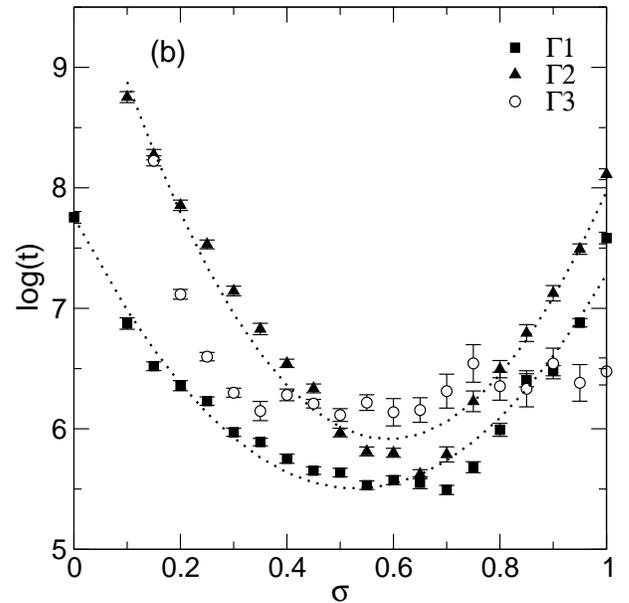}}}
{\caption{(a) Dependence of the logarithmic folding time, $\log_{10}(t)$, on the range bias parameter $\sigma$
for structures $\Gamma_{1}$, $\Gamma_{2}$ and $\Gamma_{3}$ with different 
native state's energies and with a fixed native state's energy (b).\label{fig:no4}}}
\end{figure}
The dependence of 
the folding time on the range bias parameter
is reported in Figure~\ref{fig:no4}(a) for the three native geometries. 
For $\sigma < 0.20$  (resp. $\sigma < 0.10$) we have not observed folding 
to  target $\Gamma_{3}$ (resp. $\Gamma_{2}$). \par
The behaviour exhibited by target $\Gamma_{3}$ is easily explained: since 
approximately 80 percent of $\Gamma_{3}$'s native contacts are LR there is 
little competition (and therefore little frustration) between LR and local 
interactions.
We stress that, in the present model, the competition between local and LR 
contacts results from the relative weight of the two types of interactions 
(which are always stabilizing, i.e., $<0$). The resulting frustration is 
therefore different from that of sequence-specific models where the energy 
of the local and of the LR pair interactions can be stabilizing 
(i.e., $<0$) or de-stabilizing (i.e., =0 or $>0$).
The slight decrease in the folding time observed for $\sigma >0.5$ is 
driven by the native state's energy (since its decrease with $\sigma$ 
results in a driving force to folding).
However, the effect of decreasing $\sigma$ below $\sigma=0.5$ is 
equivalent to that of 
progressively `switching-off' the LR interactions and, in the limit of 
$\sigma=0$, it actually forces the structure to fold with only 20 
per cent of its native interactions. 
In this case the driving force to folding decreases steadily  
which results in longer folding times and eventually, for $\sigma < 0.20$, 
folding failure. The observed threshold is smaller for target 
$\Gamma_{2}$ because, by contrast to the behaviour of target 
$\Gamma_{3}$, the native state's energy decreases with $\sigma$ 
(for $\sigma <0.5$) and this effect balances that of switching-off the LR 
interactions.\par      
The results obtained for the low- and intermediate-CO 
target structures, $\Gamma_{1}$ and $\Gamma_{2}$, are more interesting. 
The corresponding curves 
are qualitatively similar but a closer inspection reveals an important 
difference, 
namely: for $\sigma < 0.5$ the dependence of the folding time on $\sigma$ is 
much stronger for the intermediate-CO structure, $\Gamma_{2}$. Indeed, in this 
case one observes a remarkable three-order of magnitude dispersion of  
folding times, ranging from $\log_{10}(t_{min})=5.76 \pm 0.05$ 
(for $\sigma=0.65$) to 
$\log_{10}(t_{max})=8.75 \pm 0.05$ (for $\sigma=0.10$), by contrast with 
$\Gamma_{1}$ for which
$\log_{10}(t_{min})=5.50 \pm 0.08$ (for $\sigma=0.70$) and 
$\log_{10}(t_{max})=7.69 \pm 0.09$ (for $\sigma=0.00$). 
We note that for both structures the folding time increases considerably 
faster 
when $\sigma $ decreases than when $\sigma $ increases away from the 
minimum. However, in the latter case, the folding times do 
not deviate from each other by contrast with their behaviour for 
$\sigma < 0.5$. 
We stress that for both geometries successful folding is still 
observed for $\sigma=1.0$; this corresponds to 'switching-off' all local 
interactions, which are more than half the total number of 
interactions in both structures. 

\subsection{C. Folding kinetics for different range bias at fixed native energy}

In order to rule out differences in the folding dynamics driven by the 
stability
of the native state we now investigate the contribution of LR and local 
interactions 
to the folding kinetics of structures $\Gamma_{1}$, $\Gamma_{2}$ and 
$\Gamma_{3}$ in the following way: 
the total energy
of the native structure is kept fixed (and equal to $E=$-28.5 which is 
equivalent
to taking $\epsilon=0.5$ in Equation~\ref{eq:no1}) while the relative 
contributions of LR 
and local interactions are varied over the whole range. 
We impose the fixed energy constraint by taking the total energy 
given  
by Equation~\ref{eq:no2} and using for the long-range and local Hamiltonians
\begin{equation}
H_{LR(L)}(\lbrace \vec{r_{i}} \rbrace)=-\sum_{i>j}^N \epsilon (\sigma )\Delta_{LR(L)}
(\vec{r_{i}}-\vec{r_{j}}),
\label{eq:no4}
\end{equation}
\noindent
with 
\begin{equation}
\epsilon (\sigma ) = \frac {0.5}{(1-\sigma ) Q_L + \sigma (1 - Q_L)}
\label{eq:no5}
\end{equation}
\noindent
where $Q_{L}$ is the number of local native contacts normalised to the 
total number of native contacts. 
Again, the parameter $\sigma$ determines the contribution of local and
LR contacts to the total energy. 
For $\sigma =0.0$ ($\sigma =1.0$) only local (LR)
contacts contribute to the total energy. However, $\epsilon (\sigma)$ that 
measures 
the interaction energy of all native contacts
in $H_{LR(L)}$ varies with $\sigma $ in order to keep the 
total
energy of the conformation constant. Using Eqs.~\ref{eq:no2}, 
~\ref{eq:no4} and ~\ref{eq:no5} the energy per native contact
is given by $\epsilon_{L}=(1-\sigma)\epsilon(\sigma)$ if the contact is 
local while it is given by 
$\epsilon_{LR}=\sigma \epsilon (\sigma)$ for LR contacts. Figure 
~\ref{fig:no5} shows the
dependence of $\epsilon_{LR}$ and $\epsilon_{L}$ on the range bias parameter 
for 
structures
$\Gamma_{1}$, $\Gamma_{2}$ and $\Gamma_{3}$.\par

We have studied the equilibrium population of states in order to investigate 
the native state's occupation probability at the optimal folding temperature 
as well as its dependence on $\sigma$. To this end 
long simulations (lasting up to $10^{8}$ MC steps) were preformed in order to 
ensure that data was collected under equilibrium conditions~\cite{GUTIN}. 
The results from these simulations (for the three targets) are 
reported in the histograms of Figure~\ref{fig:no6} for values of 
$\sigma=0.3,0.5,0.7$. The height of each bar in the histograms, measures the 
probability occupancy, i.e., the number of molecules (normalised to the total 
number of molecules collected in one run) with a fraction of native contacts 
$Q$. In all the cases considered most molecules are
in the native state ($Q=1.0$). However, when $\sigma=0.3$, the native state 
of target $\Gamma_{1}$, has a rather low probability occupancy (less than 0.5).
As the stability of the native state is estimated as being proportional to 
the probability of the chain to be in the native 
conformation~\cite{GUTIN}, we conclude that, for $\sigma=0.3$,
the native state of target $\Gamma_{1}$ is not very stable. We note that
target $\Gamma_{3}$, which has the largest fraction of long-range contacts, 
exhibits the highest native state ocupation probability for all the three 
values 
of $\sigma$. This observation is in line with the idea that long-range 
contacts have a dominant role in stabilizing the native fold.\par
Figure~\ref{fig:no4}(b) shows the dependence of the folding time, $t$,
on the range bias parameter for the three targets at fixed native state's 
energy. The conclusions drawn for the case of varying native state's energy 
hold equally well in the fixed energy case.  
In particular, the reported results confirm the  
trend for the dependence of $\Gamma_{3}$'s folding time on the range bias 
parameter. We should stress, however, that 
in the present case folding failure is observed for $\sigma < 0.15$ by 
contrast with the varying energy model where no successful folding
was observed for $\sigma <0.20$. We ascribe this behaviour to
the stabilizing (or equivalently, to the lower) native state's energy 
which compensates the
effect of `switching-off' the LR interactions. 
Hereafter we will restrict the discussion to the results for 
structures $\Gamma_{1}$  and $\Gamma_{2}$.
We note that the main difference between the fixed and varying energy 
models is that for $\sigma >0.5$ the 
folding times are systematically longer 
(up to one order of magnitude for $\Gamma_{2}$) 
when the native state's energy is kept fixed. 
\begin{figure} 
\centerline{\rotatebox{0}{\resizebox{8cm}{8cm}{\includegraphics
{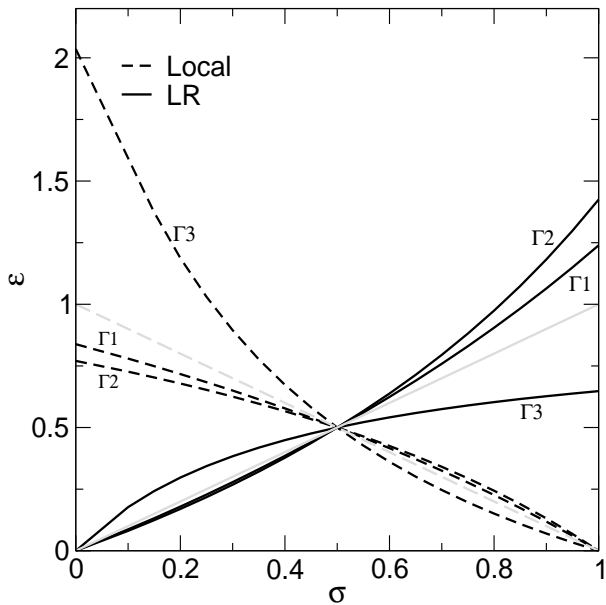}}}} 
\caption{Dependence of the $\epsilon_{LR}$ (energy per long range contact) and 
$\epsilon_{L}$  (energy per local contact) on the
range bias parameter $\sigma$ for structures $\Gamma_{1}$, $\Gamma_{2}$ and 
$\Gamma_{3}$ in the fixed energy model. Also shown (in light grey) the dependence of 
$\epsilon_{LR}$ and $\epsilon_{L}$ on $\sigma$ in the case of the varying native energy model.\label{fig:no5}}
\end{figure}
Recall from Figure~\ref{fig:no3} that when the native state's energy is 
allowed to vary it 
increases with $\sigma$ up to $E=-23$ and $E=-20$ for targets $\Gamma_{1}$
and $\Gamma_{2}$, respectively. The 
fixed native state's 
energy $E=-28.5$ is lower than the varying native energies in the range 
$\sigma > 0.5$.
Should native energy play a significant role, the folding time's dependence on 
$\sigma$ for $\sigma > 0.5$ would be less pronounced for the fixed native 
energy simulations, in sharp contrast with our findings. Instead, these 
are consistent 
with the idea that the kinetics of folding is dominated by the formation of 
LR contacts.
As shown in Figure~\ref{fig:no5}, in $\Gamma_{1}$ and $\Gamma_{2}$ the energy
bias favouring LR contacts for $\sigma > 0.5$ is greater in the fixed energy
case. This explains the differences in the behaviour of the curves 
corresponding to these two targets in Figures~\ref{fig:no4}(a) and 
~\ref{fig:no4}(b).\par
From the results reported in Figure~\ref{fig:no4}, we conclude that, 
by comparison with local contacts, LR contacts play
a crucial role, in driving the folding kinetics of small G\={o}-type lattice 
polymers. 
Moreover, the effect of LR contacts on the kinetics is strongly dependent on 
the native state's geometry.

\subsection{D. Native structure and the geometric coupling between long-range and local contacts}

In this section we investigate the differences between the folding 
processes of
targets $\Gamma_{1}$ and $\Gamma_{2}$ when $\sigma$ is varied from zero
to one in order to interpret the behaviour observed in the previous section. \par
Recall from section II that our `reaction' coordinate is the fraction of 
native contacts $Q$. In general, $Q$ works as a thermodynamic reaction
coordinate by measuring closeness to the native structure in energetic terms
only. As argued in Ref.{\cite{CHAN}} thermodynamic closeness does not 
necessarily
imply kinetic proximity to the native structure unless the energy landscape
is considerably smooth {\cite{JEWETT}} (as it happens to be the case in 
the present study). 
Indeed under such conditions one can take $Q$ as a kinetic
reaction coordinate so that it actually defines how quickly a conformation
can convert into the native structure {\cite{CHAN}}. Thus in what follows 
we assume that $Q$ measures the kinetic progress towards the native state. \par
In Figure~\ref{fig:no7} we have plotted for targets $\Gamma_{1}$ and $\Gamma_{2}$, 
and for three values of $\sigma$ (namely, 0.5, 0.1 and 1.0),
the dependence on $Q$ of the following kinetic quantities: the fraction of 
LR contacts,
$q_{LR}$, the fraction of local contacts, $q_{L}$ and the normalized 
logarithmic folding
time, $log_{10}(t^{*})$ (note that in this case the fractions of LR and 
local contacts are normalized to the
total number of LR and local native contacts, respectively, and not to the 
total number of native contacts; therefore $q=1.0$ when $Q=1.0$ but in general $q 
\ne Q$ and this is why the adopted notation is different from that used in the 
previous sections.\par 
We start by observing the unbiased case, that is $\sigma=0.5$. The kinetics of 
local contact formation is similar in both targets with 
the fraction of local contacts starting from a much larger value than that 
of LR ones. However, long-range
contacts form considerably earlier in $\Gamma_{1}$ and, in this case,
the kinetics of LR contacts follows that of $q_{L}$. In both 
cases, the normalized folding time 
is controlled by the formation of local contacts.  
For $\sigma=1.0$ local interactions are switched-off and this results in 
an effective slowing-down of 
the corresponding kinetics in both targets. Note that, in 
$\Gamma_{1}$, $q_{LR}$ grows 
extremely quickly reaching $\approx$ 90 per cent relatively early in 
the folding process (when $Q=0.58$) when compared with
the behaviour exhibited for $\sigma=0.5$ (90 per cent when $Q=0.92$). 
However, this early boost in $q_{LR}$
does not lead to stable structure formation as it subsequently 
drops-down to $q_{LR}=0.84$ (for $Q=0.86$) and is then forced to follow 
the kinetics of local contact 
formation. For this value of $\sigma$, the
folding time is controlled by the setting-up of LR contacts in 
$\Gamma_{2}$ and by that of local
contacts in $\Gamma_{1}$. Finally, when $\sigma=0.1$, both targets 
exhibit a similar dependence of $q_{L}$ 
on $Q$ but the kinetics of $q_{LR}$ is slowed-down considerably in 
$\Gamma_{2}$.
Again, the folding time is controlled by $q_{L}$ in $\Gamma_{1}$ and
by $q_{LR}$ in $\Gamma_{2}$, but for $\Gamma_{2}$, the setting-up of LR 
contacts is much slower than in the previous cases.\par
We interpret these observations in the following way: in target 
$\Gamma_{1}$ there is a strong 
geometric coupling between the formation of local and LR contacts, meaning 
that the establishment of LR
contacts is promoted by the establishment of local contacts. On the 
other hand, in target
$\Gamma_{2}$, there is no such coupling and this results in an overall 
kinetics which is more sensitive
to changes in the energy interaction parameters. In particular, in  
$\Gamma_{2}$, local 
contacts are not capable of promoting the setting-up of LR contacts when 
the LR interactions are 
highly penalized energetically (recall that we did not observe 
successful folding to $\Gamma_{2}$ when $\sigma < 0.10$).

\section{IV. Conclusions and final remarks}

In this paper we have revisited the G\={o} model in order to investigate, by
means of a new approach, the role of long-range(LR) and local interactions 
in the folding kinetics.
We have focused our analysis on lattice-polymers with chain length $N=48$ 
since, like real two-state folders, these models exhibit relatively 
smooth energy 
landscapes. We studied the changes in the folding process induced by 
unbalancing the contributions of local and LR interactions to the
native state's energy.
Our results strongly suggest that LR interactions play a dominant
role in the folding kinetics. Indeed we observe a
decrease in the folding rates, or equivalently, an increase in the 
folding time, which is clearly more pronounced when the contribution 
of the LR interactions (relative to 
that of the local interactions) to the native state's energy  
is progressively decreased towards zero. 
We have found that this effect is essentially independent of the native 
state's energy.
By contrast, the kinetic response to decreasing the relative 
contribution of the LR interactions is strongly dependent on the target 
geometry. We 
have selected our target geometries on the basis
of differing contact order parameters. In the high-CO target, 
$\Gamma_{3}$, LR contacts 
are the vast majority (44 out of
57 native contacts) and this results in a trivial kinetic response: when 
LR interactions are
strengthened relative to the local ones there are no significant changes 
in the folding rates;
on the other hand, a strong increase in the folding rates (eventually 
resulting in folding failure) arises 
when they are weakened.
Interesting behaviour occurs in the folding kinetics of the other two 
structures, 
the low- and intermediate-CO targets,
$\Gamma_{1}$ and $\Gamma_{2}$, respectively. In both structures local 
contacts dominate and both exhibit 
a similar fraction of local and LR contacts.
However, in the intermediate-CO target the kinetics is much more sensitive 
to the weakening of LR interactions; 
in fact in this case we observed a remarkable three-order of 
magnitude 
dispersion of folding times, although this is still two-orders of 
magnitude smaller than the dispersion of folding times of real two-state 
folders 
($\approx$ 5 orders of magnitude) {\cite{PLAXCO2}}. \par

The topomer search model (TSM) for protein folding 
(reviewed in~\cite{PLAXCO3}) considers that the folding time is determined 
by the difusive search for the ensemble of unfolded structures that share 
a similar, global topology with the native state (the native 
topomers)~\cite{DEBE}. Achieving the native topomer corresponds to 
surmounting the rate limiting step in folding, which is followed by the 
zippering of specific local native contacts, a process that rapidly leads to 
the native structure.
Thus, according to the TSM, the rate at which an unfolded protein diffuses 
between distinct topologies is much slower than the rate at which local 
structural elements form. Recent results obtained through numerical 
simulations of the diffusion of Gaussian chains by Makarov {\it et 
al.}~\cite{PLAXCO3,MAKAROV} suggest, in the context of this model, that the 
logarithmic folding rate grows linearly with the number $N_{LR}$ of LR 
contact pairs in the native structure, which define the topomer.    
Makarov {\it et al.}~\cite{PLAXCO3} investigated wether the TSM 
correlates
well with the folding rates of the 24 two-state folders previously studied 
by Plaxco {\it et al.}~\cite{PLAXCO2}. To determine $N_{LR}$ for each protein
the authors have considered as LR a native contact where the amino acids 
are separated by at least 12 or more residues along the protein backbone. 
A considerably strong correlation ($r=0.88$) was found between the 
logarithmic folding rates and $N_{LR}$~\cite{PLAXCO3}, suggesting that the 
TSM is indeed a plausible model for two-state folding rates. Our 
results are in broad agreement
with the TSM in the sense that, irrespectively of target geometry, we have 
found that decreasing, versus increasing, the relative weigth of LR 
interactions leads to a more pronounced increase
of the folding times. However, we have also found evidence for a folding 
mechanism (on the lattice) that is different from that of the TSM. 
According 
to the TSM, the step that determines the folding rates is the formation of 
the LR contacts in the
native topomer. After this step a rapid zippering of the local contacts 
occurs and the
native structure forms. Our results show that, depending on native geometry, 
the formation of LR contacts may be more strongly coupled with the formation 
of the local contacts. This is clearly illustrated by the behaviour of 
target
$\Gamma_{1}$ when $\sigma=0.1$ (Figure~\ref{fig:no7}). For this value of
$\sigma$ the LR contacts are strongly penalized energetically, 
and
the folding time is controlled by local contact formation. This is not 
observed for target $\Gamma_{2}$, where under the same conditions 
($\sigma=0.1$) the folding time is controlled by LR contact formation 
(Figure~\ref{fig:no7}), in agreement with the TSM.
Another result that supports the existence of coupling between local and 
LR
contact formation in target $\Gamma_{1}$ is the fact that, again for 
$\sigma=0.1$, LR contact formation is much faster for target $\Gamma_{1}$ 
than for target $\Gamma_{2}$.
We note that this particular aspect
of the folding process in lattice models has not been explored in previous 
simulation efforts.
\par
In a recent study Micheletti {\it et al.}~\cite{MARITAN1} have introduce a 
novel method,
the so-called `geometrical variational principle', to investigate the role 
of native geometry 
in guiding the protein to the native fold. This study consisted in computing
the number of structures that share a certain structural similarity with a 
given native structure (the structural similarity between a structure and 
the fixed native fold
is measured by the fraction of native contacts $Q$ in that structure). The 
authors have called this measure the density of overlapping conformations 
(DOC). A crucial result from this study was the finding that the DOC of 
real natural folds is always much larger (at any value of $Q$) than that found 
in artificially generated structures (with the same chain length and 
number of contacts but differing in the fractions of local and non-local 
contacts). Moreover, the authors found that, for $Q \approx 0.5$, the 
DOC of real folds is very close to its maximum value and that this 
`extremality' of the DCO is related with a high content in 
secondary-structure-like motifs (alpha-helices and beta-sheets). In a 
subsequent study Maritan {\it el al.}~\cite{MARITAN2} applied a 
`dynamical variational principle' (DVP) to search for rapid folders in 
conformational space. 
The authors have found, in the context of a G\={o} model on a fcc lattice, 
that decreasing folding times are associated with increasing secondary 
structure content (in agreement with Micheletti's results) and with 
decreasing contact order. This finding shows that the aplication of the 
DVP to search for kinetically foldable proteins results in the emergence 
of structures with predominantly helical order (i.e., with a high content 
in local contacts). Since our results were obtained for a cubic lattice a 
detailed comparison with Maritan's findings is not possible. However, in  
the contact map of 
Figure~\ref{fig:no1}(a), corresponding to $\Gamma_{1}$, one can clearly 
identify a pattern that
resembles that of alpha-helices namely, the existence of thick bands 
parallel
to the main diagonal. The fact that $\Gamma_{1}$ exhibits the 
shortest folding times for
all values of the LR interaction strength is in broad agreement with 
Maritan's results.\par 
The existence of a geometric coupling between local and LR contacts
may have implications in what concerns the understanding of protein 
evolution
in the sense that it provides a possible mechanism for the emergence of 
mutational robustness in proteins. A biological system is said to be robust 
to mutations if it continues to function after genetic changes in any 
of its parts. 
Native structures endowed with a mechanism of local-LR contact coupling are 
naturally more capable of exhibiting a fast adaptation to mechanisms
of biomolecular variation (point mutations, insertions, deletions, etc) that 
change the amino acid sequence (i.e. that change the set of amino acid 
interactions) in the following sense. If the geometric coupling between 
local and LR contacts exists, one expects the protein's foldability (the 
protein's ability to fold at a reasonably fast rate which is indeed an 
evolutionary advantage) to be less affected by changes in the way the amino 
acids interact since when the LR contacts become energetically penalized, 
as a result of sequence changes, the establishment of local contacts acts 
as a driving force for the establisment of LR ones.\par
Finally, it would be interesting to investigate the interplay between target geometry
and favored native contact interactions in more realistic models, where not only
the dispersion of folding times of real proteins is reproduced as well as
other aspects observed in the folding of real two-state folders such as the 
thermodynamic and the kinetic cooperativities{~\cite{ENZYMOLOGY}}. A simple
model that is a step in this diection is that of Kaya and Chan who used  
a modified G\={o}-type potential, involving nonadditive multybody interactions, 
to study the folding dynamics of 27-mers on a cubic lattice~\cite{KAYA}.
When applied to a pool of targets comprising 97 native geometries, chosen on the basis of their CO parameters, Kaya and Chan's model yielded folding rates spanning more than 2.5 orders of magnitude. Furthermore this model also exhibited thermodynamic cooperativity
and linear chevron plots (i.e., kinetic cooperativity) similar to those observed in experiments with real proteins.

\begin{figure*}[!ht] 
{\rotatebox{270}{\resizebox{5.5cm}{5.5cm}{\includegraphics {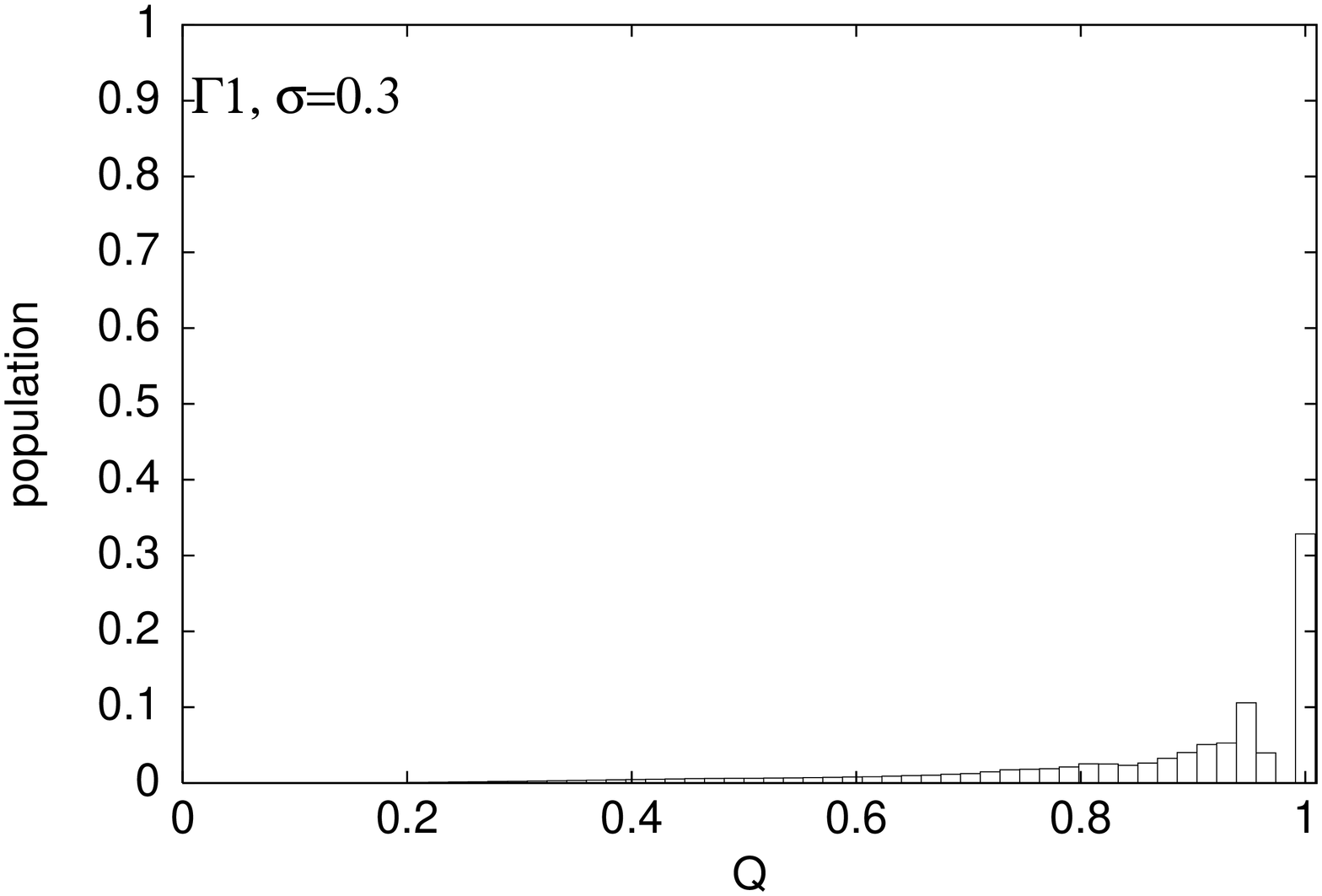}}}} \hspace{0.5cm}
{\rotatebox{270}{\resizebox{5.5cm}{5.5cm}{\includegraphics {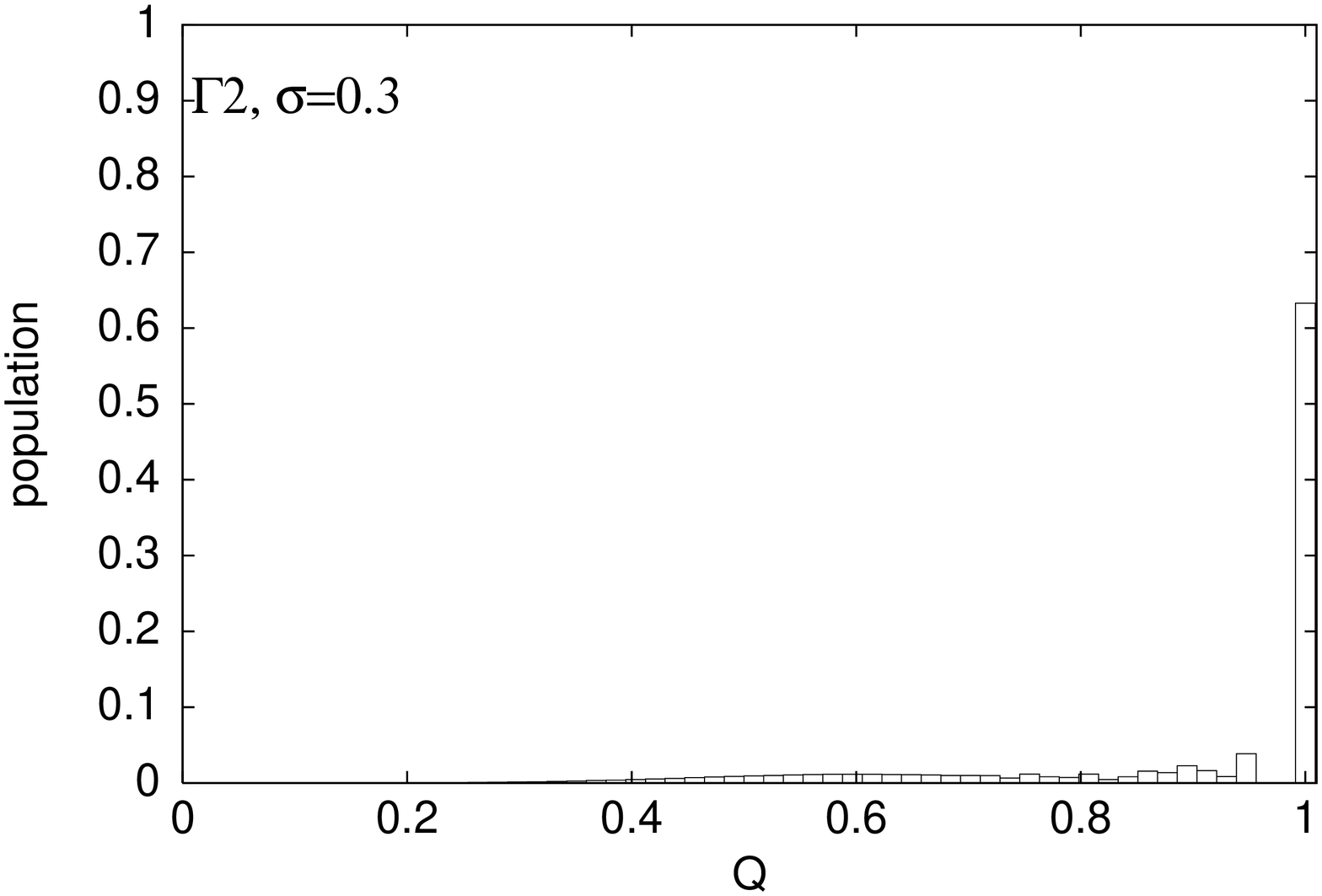}}}} \\
\vspace{0.5cm}
{\rotatebox{270}{\resizebox{5.5cm}{5.5cm}{\includegraphics{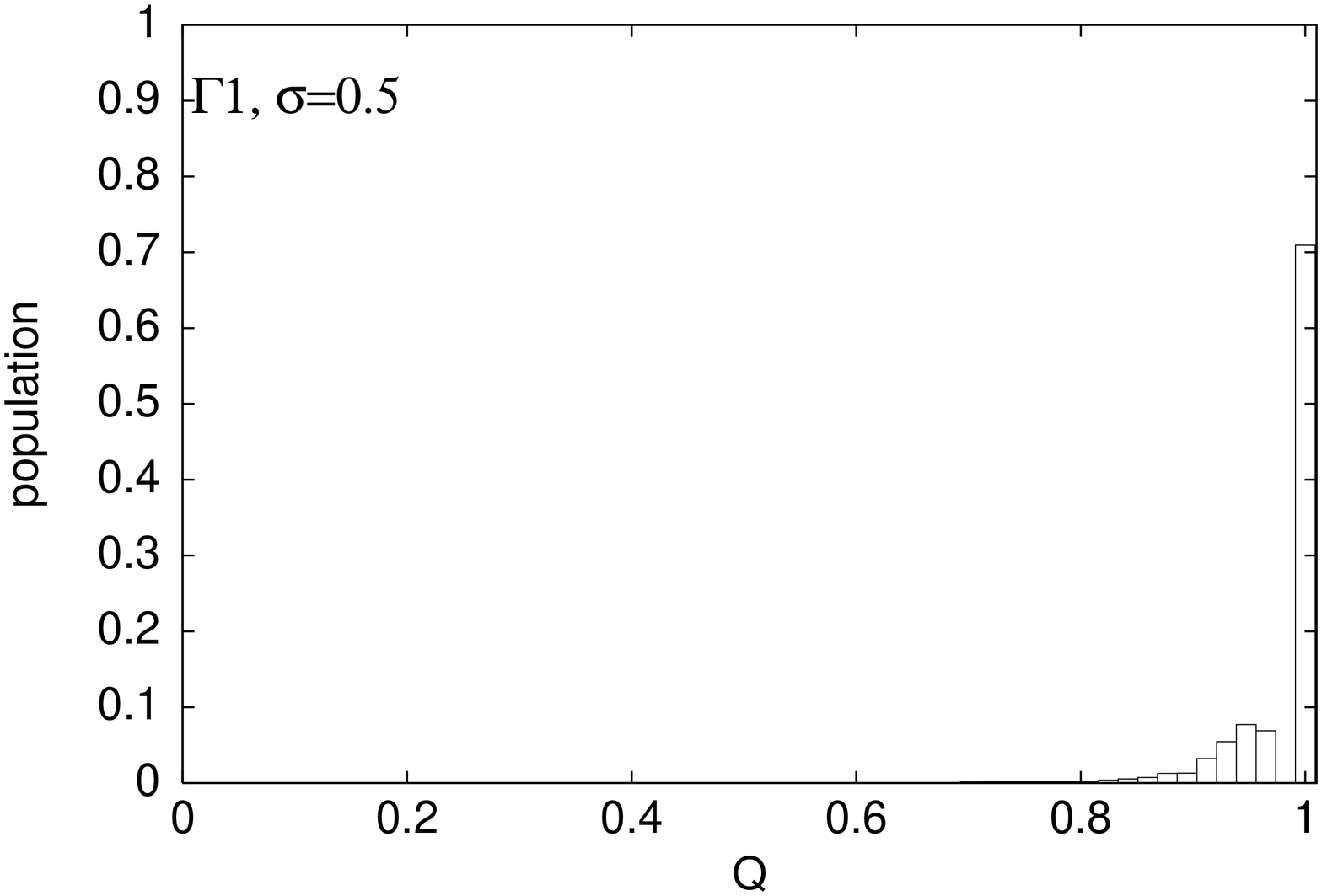}}}} \hspace{0.5cm}
{\rotatebox{270}{\resizebox{5.5cm}{5.5cm}{\includegraphics{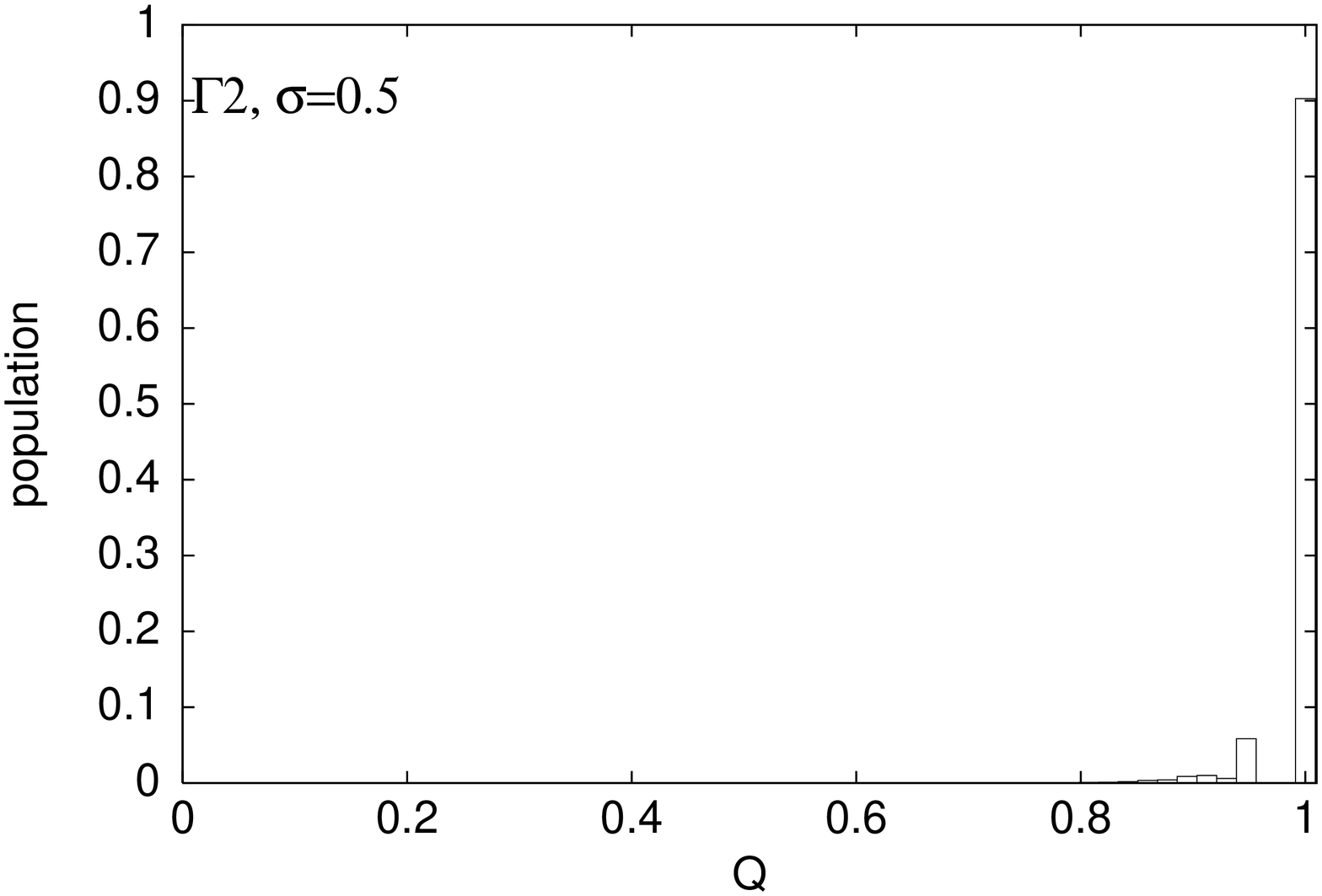}}}} \\
\vspace{1cm}
{\rotatebox{270}{\resizebox{5.5cm}{5.5cm}{\includegraphics{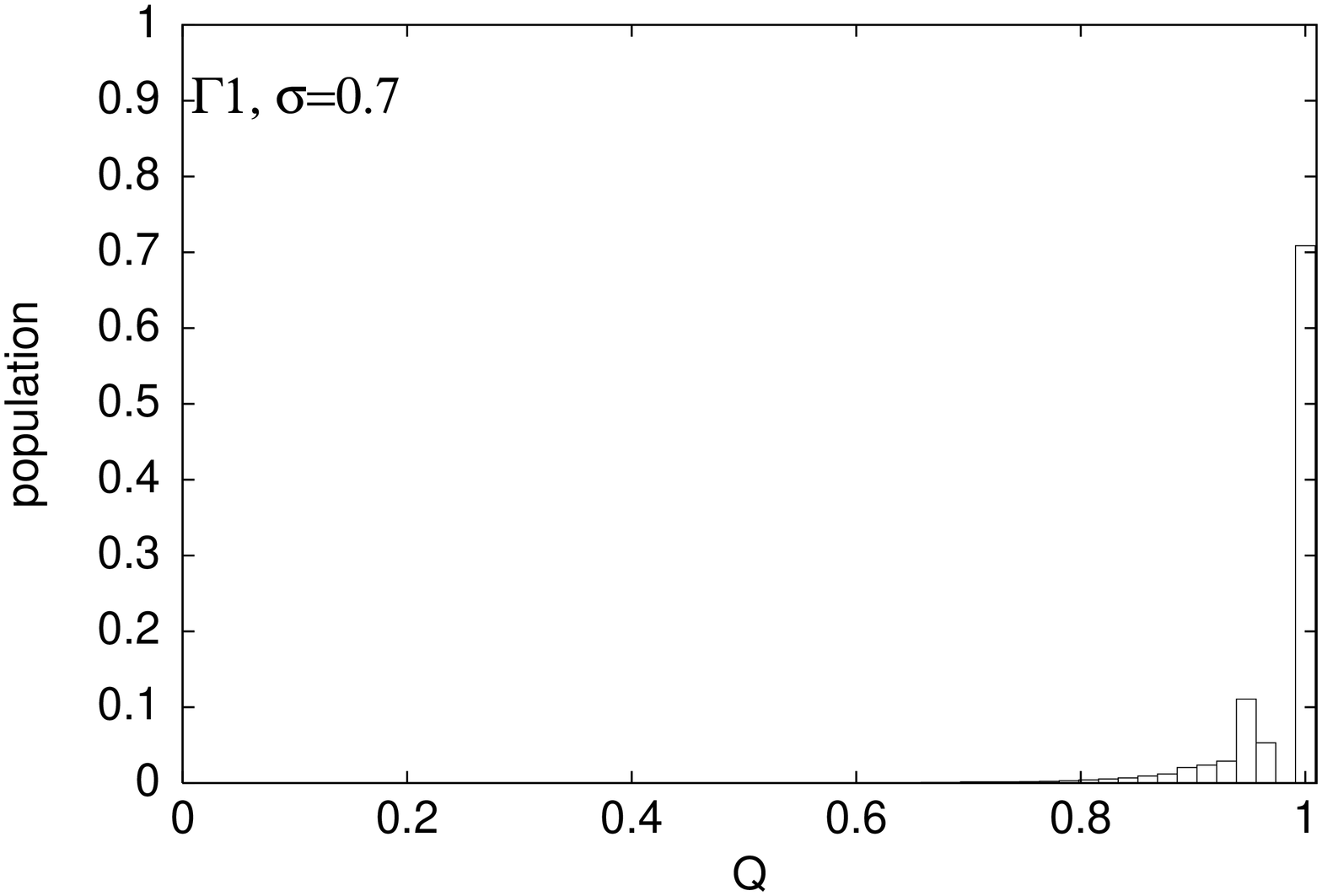}}}} \hspace{0.5cm}
{\rotatebox{270}{\resizebox{5.5cm}{5.5cm}{\includegraphics{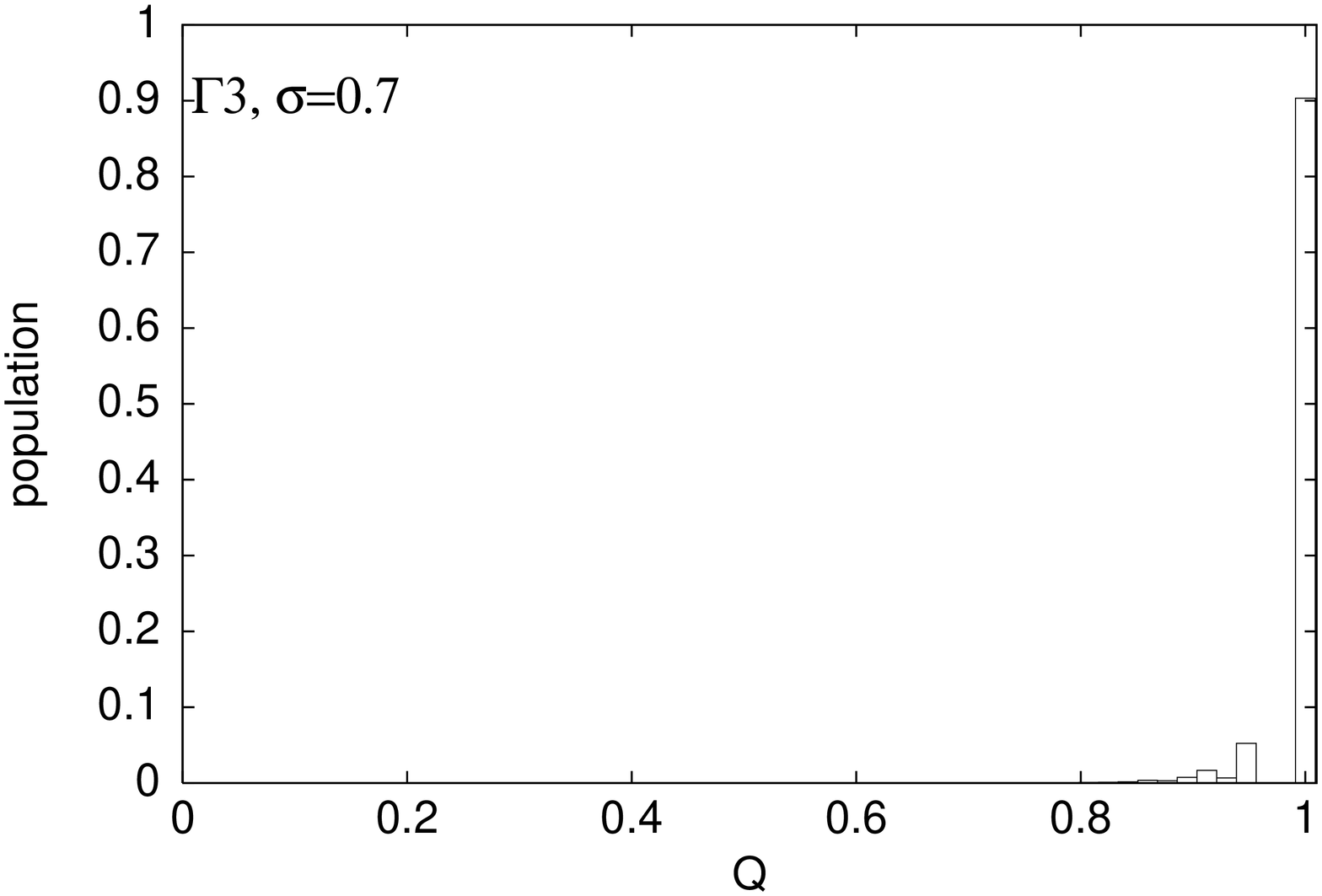}}}} 
\caption{Population histograms, for targets $\Gamma_{1}$ (first row),  
$\Gamma_{2}$ (second row) and $\Gamma_{3}$ (third row) and 
$\sigma=$0.3, 0.5 and 0.7 at the optimal folding temperature. 
The native state, corresponding to $Q=1.0$, is the dominant state for 
all structures at all values of $\sigma$. Except for structure 
$\Gamma_{1}$ at $\sigma=0.3$ the native state's occupation 
probability is larger than 0.5. Target $\Gamma_{3}$, with the largest 
fraction of long-range contacts, exhibits the largest native state 
ocupation 
probability at all values of $\sigma$. This observation is in line 
with the idea that long-range contacts have a dominant role in stabilizing 
the native fold.
\label{fig:no6}}
\end{figure*}

\begin{figure*}[!ht] 
{\rotatebox{0}{\resizebox{5.5cm}{5.5cm}{\includegraphics
{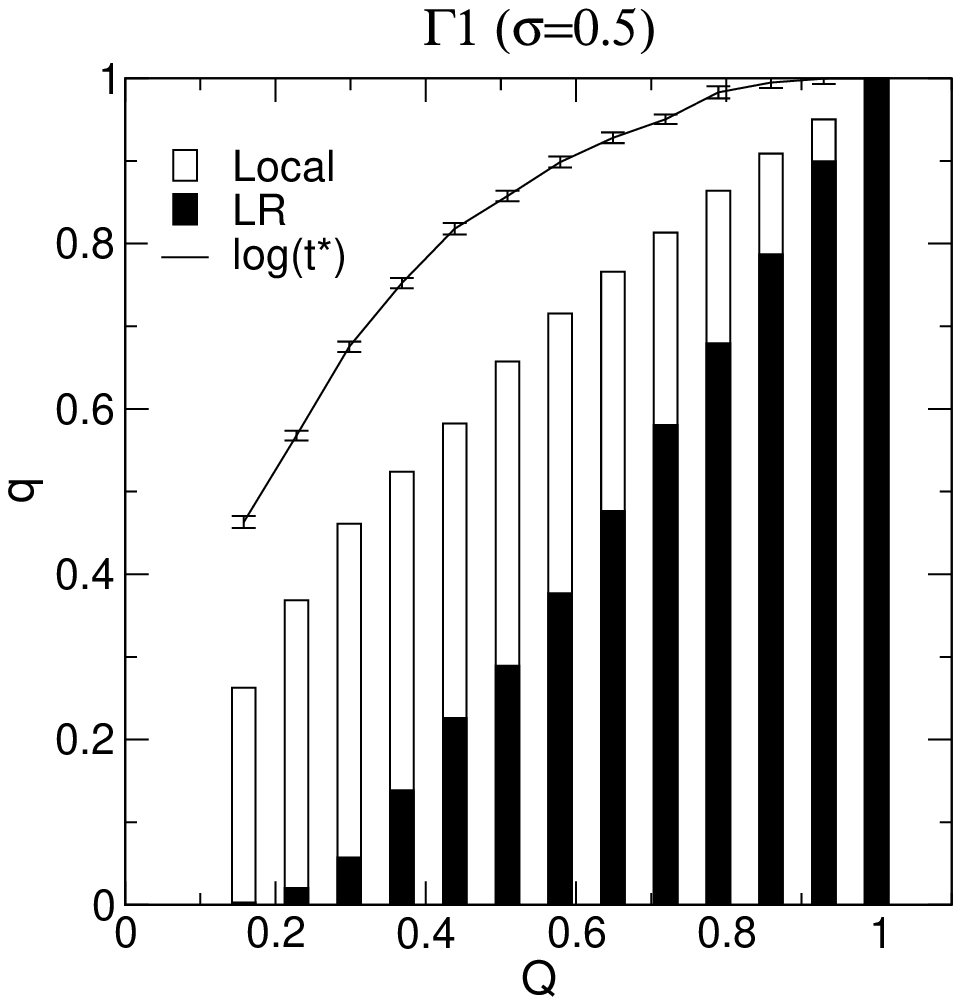}}}} \hspace{0.5cm}
{\rotatebox{0}{\resizebox{5.5cm}{5.5cm}{\includegraphics
{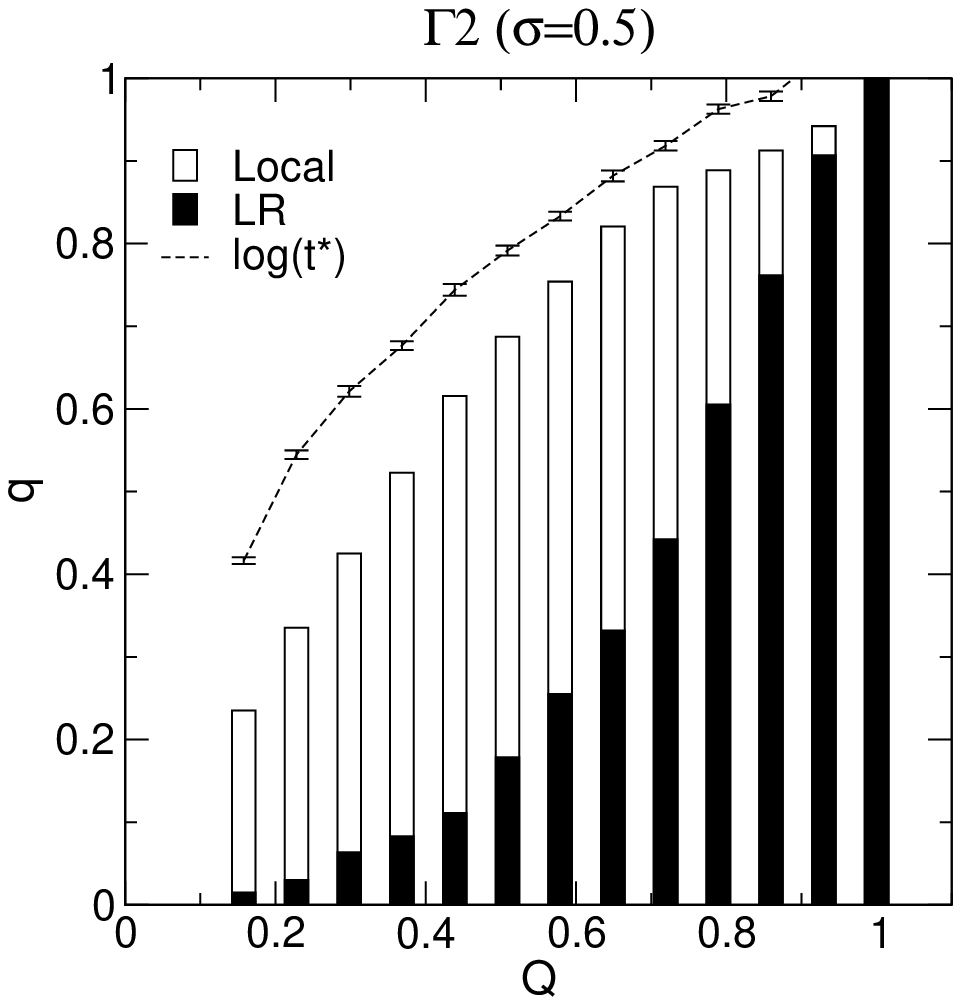}}}} \\
\vspace{0.5cm}
{\rotatebox{0}{\resizebox{5.5cm}{5.5cm}{\includegraphics
{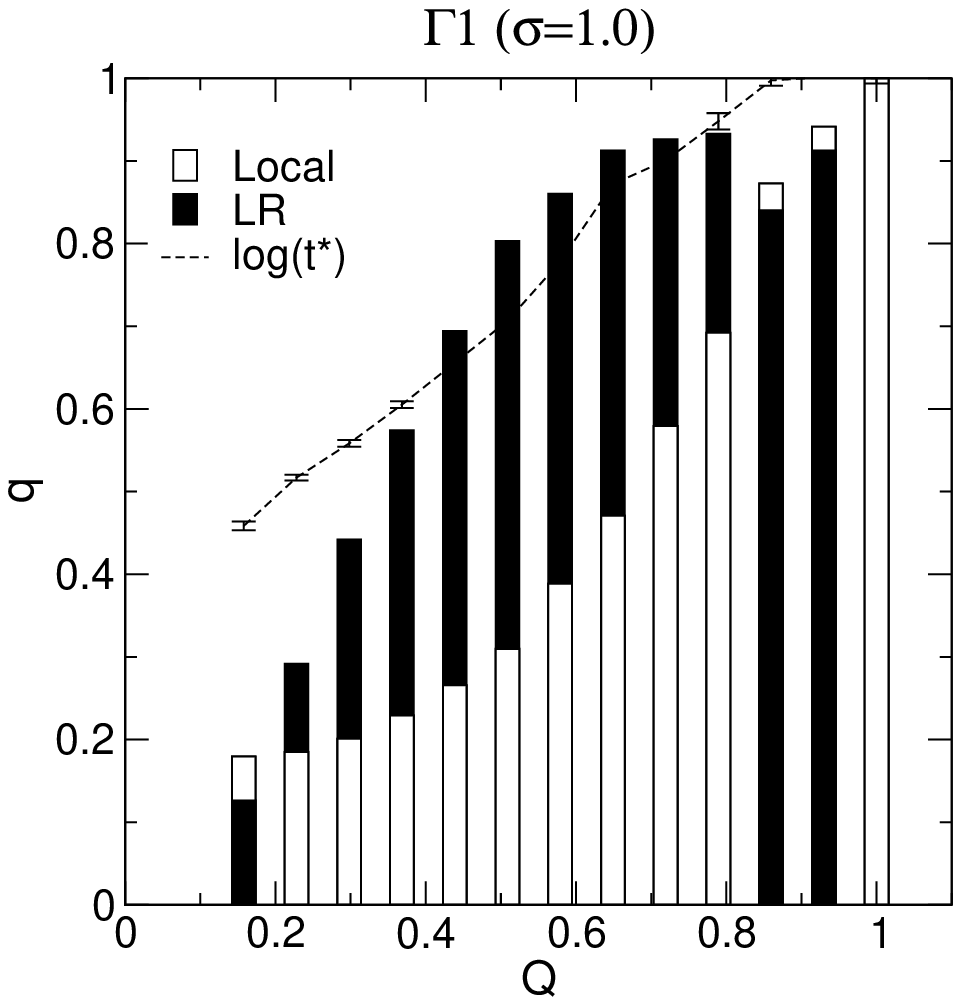}}}} \hspace{0.5cm}
{\rotatebox{0}{\resizebox{5.5cm}{5.5cm}{\includegraphics
{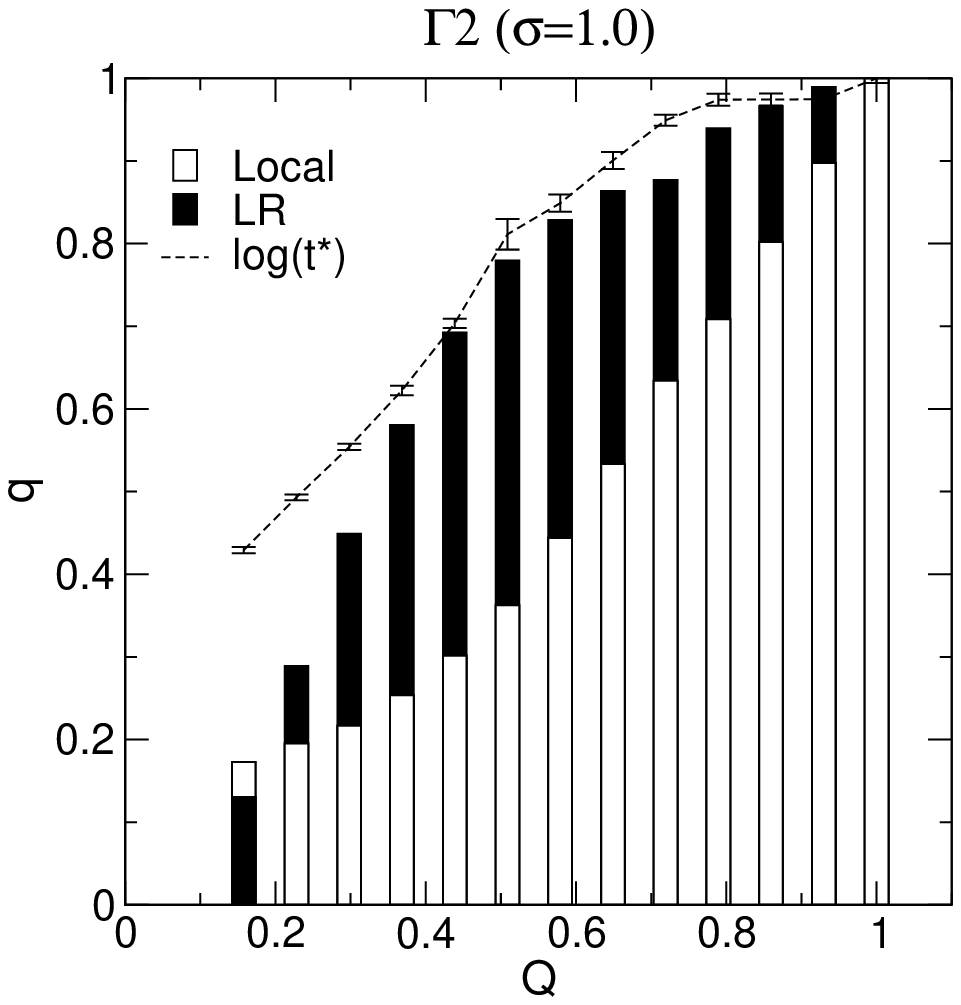}}}} \\
\vspace{1cm}
{\rotatebox{0}{\resizebox{5.5cm}{5.5cm}{\includegraphics
{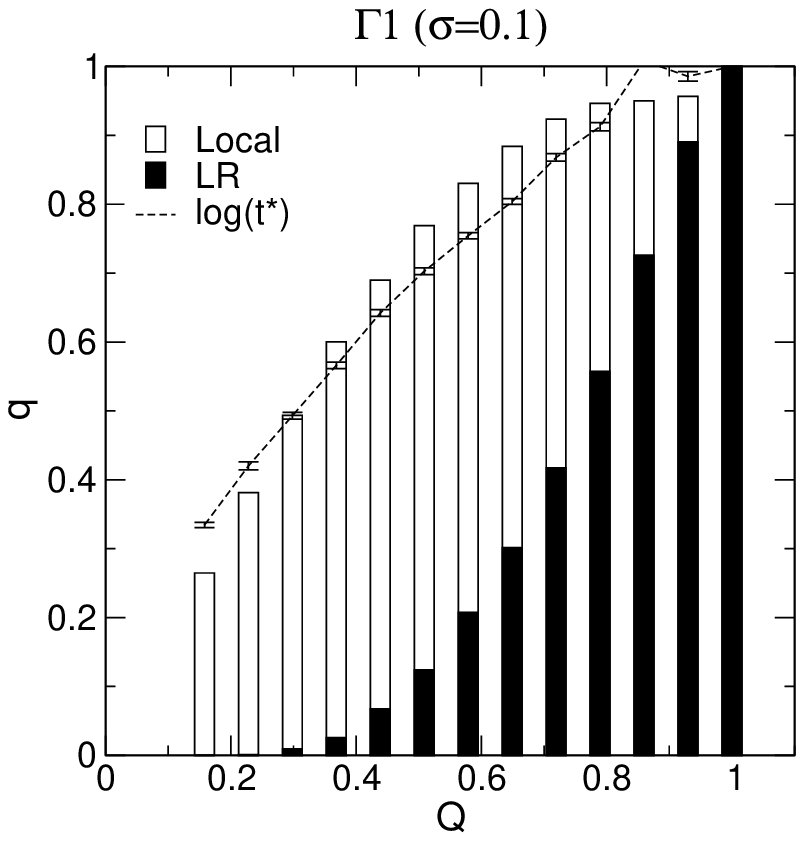}}}} \hspace{0.5cm}
{\rotatebox{0}{\resizebox{5.5cm}{5.5cm}{\includegraphics
{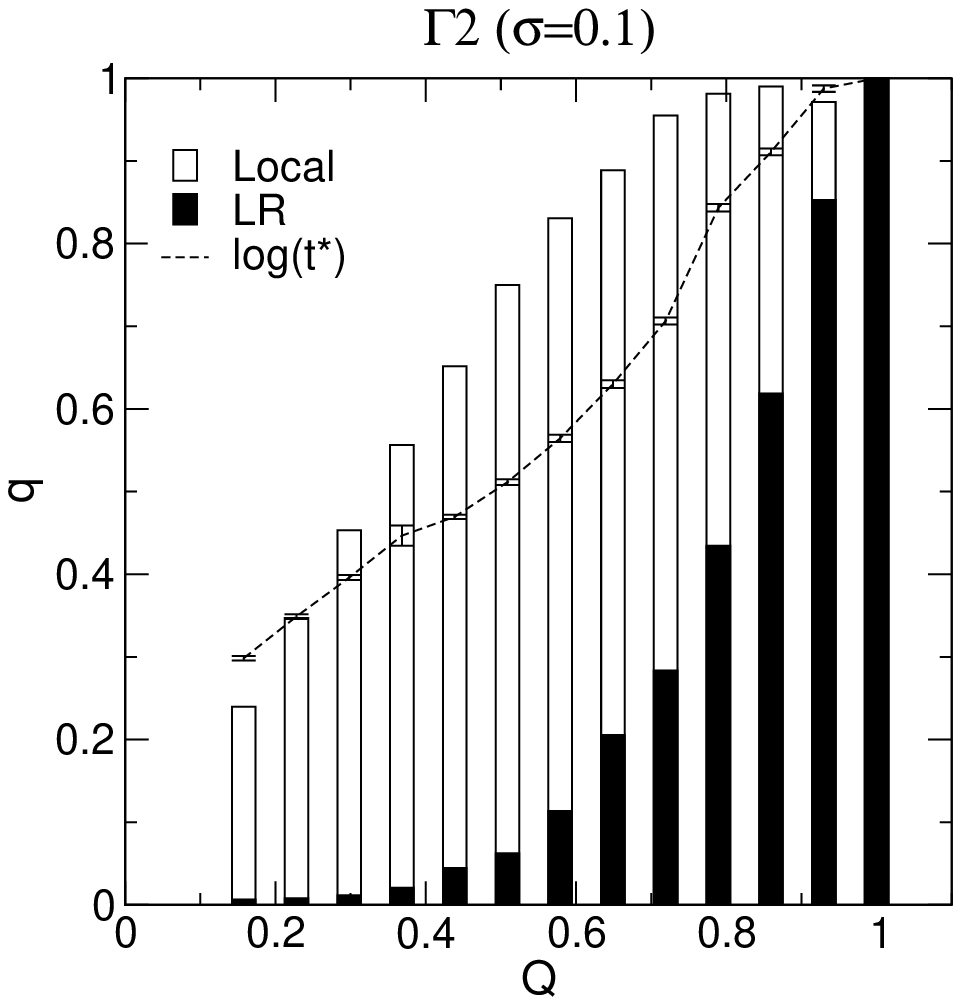}}}} 
\caption{Dependence of the fractions of long-range (LR) and local contacts 
on the fraction of native contacts, $Q$, for targets $\Gamma_{1}$ and 
$\Gamma_{2}$, and $\sigma=0.5, 1.0$ and 0.1. 
Note that $q$ is $q_{LR}$ (i.e., the fraction of LR contacts) for the black 
bars and $q_{L}$ (i.e., the fraction of local contacts) for the white bars. 
$q_{LR}$ ($q_{L}$) is the number of LR (local) contacts normalised to the 
total number of LR (local) contacts in each native structure. Also shown is 
the dependence of $\log_{10}(t^{*})$ on $Q$ where $t^{*}$ is folding time 
at each $Q$ normalised to the folding time at $Q=1.0$.   
\label{fig:no7}}
\end{figure*}

{\small This is a preprint of an article accepted for publication in Proteins: Structure, Function and Bioinformatics.
Copyright(1999-2005) John Wiley \& Sons. All rights reserved}

\end{document}